%% file: main.tex
\documentclass[lettersize,journal]{IEEEtran}
\usepackage[utf8]{inputenc}

\usepackage{color}
\usepackage{hyperref}

\usepackage{tikz}
\usetikzlibrary{automata,arrows,calc,math,intersections,positioning,patterns,external,tikzmark,arrows.meta}

\usepackage{pgfplots}
\usepgfplotslibrary{external,statistics,fillbetween}
\pgfplotsset{compat=1.16}

\usepackage{amsmath,amssymb,amsfonts,amsthm}
\usepackage{caption}
\usepackage{currfile}
\usepackage{algorithm}
\usepackage{algpseudocode}
\usepackage{listings}
\usepackage{siunitx}
\usepackage{graphics}
\usepackage{mathtools}
\usepackage{bytefield}
\usepackage{subfig}
\usepackage{makecell}
\usepackage{multirow}
\usepackage{enumitem}
\usepackage{adjustbox}

\title{Attacking (and defending) the Maritime Radar System}
\author
{
    Giacomo Longo, Enrico Russo, Alessandro Armando, and Alessio Merlo 
    \thanks{
        G. Longo, E. Russo, A. Armando, and A. Merlo are with University of Genoa
    }
}

\DeclareMathOperator{\atantwo}{atan2}

\newcounter{example}
\newenvironment{example}[1][]{\refstepcounter{example}\par\medskip\noindent
   \textbf{Example~\theexample. #1} \rmfamily}{\qed\medskip}

\begin{document}

\maketitle
\input{abstract}


\input{sections/1-introduction/section}

\input{sections/2-background/section}

\input{sections/3-techniques/section}

\input{sections/4-attack-description/section}
\input{sections/5-attacks/section}

\input{sections/6-detection/section}

\input{sections/7-evaluation/section}

\input{sections/8-conclusion/section}

\section{Acknowledgments}
This work was partially supported by research funding from Fincantieri S.p.A. and a grant from the National Ph.D. Programme in Artificial Intelligence (Security and cybersecurity area).
The research activities are carried out thank to the ShIL (Ship-In-the-Loop) research infrastructure, co-funded by Regione Liguria, University of Genova and DLTM under the program POR FESR LIGURIA 2014-2020 ASSE 1 "Research and Innovation (OT1)" Action 1.5.1 Notice "Support for research infrastructures considered critical / crucial for regional systems".

\bibliographystyle{IEEEtran}
\bibliography{main}

\appendices

\end{document}

%% file: abstract.tex
\begin{abstract}
Operation of radar equipment is one of the key facilities used by navigators to gather situational awareness about their surroundings.
With an ever increasing need for always-running logistics and tighter shipping schedules, operators are relying more and more on computerized instruments and their indications.
As a result, modern ships have become a complex cyber-physical system in which sensors and computers constantly communicate and coordinate.

In this work, we discuss novel threats related to the \textsl{radar system}, which is one of the most security-sensitive component on a ship.
In detail, we first discuss some new attacks capable of compromising the integrity of data displayed on a radar system, with potentially catastrophic impacts on the crew' situational awareness or even safety itself.
Then, we present a detection system aimed at highlighting anomalies in the radar video feed, requiring no modifications to the target ship configuration.
Finally, we stimulate our detection system by performing the attacks inside of a simulated environment.
The experimental results clearly indicate that the attacks are feasible, rather easy to carry out, and hard-to-detect. Moreover, they prove that the proposed detection technique is effective.
\end{abstract}

%% file: sections/1-introduction/section.tex
\section{Introduction}

\IEEEPARstart{C}{onduction} of a vessel is increasingly relying on both Information Technology (IT) and Operational Technology (OT).
The advantages brought by their adoption cannot be overestimated as OT, through the automation of onboard operations associated with the mechanical and electrical subsystems, enables a reduction of costs as well as the execution of risky tasks by the crew and IT, and more generally Information and Communication Technologies (ICT), provide invaluable support to navigation planning, control and monitoring.
As a matter of fact, commercial ships undertaking international voyages are subject to multilateral treaties mandating the installation of a variety of electronic devices~\cite{solas2020}. 
Such provisions, combined with initiatives promoted by the International Maritime Organization (IMO), e.g.~e-navigation~\cite{eNavigation,eNavigationImplementation}, have led to a significant onboard system digitization.

The Integrated Navigation System (INS) lies at the core of this digitization.
By gathering information and integrating functions from a variety of electronic devices (e.g.~the radar), the INS helps the operator to plan, monitor, and control the navigation and contributes to improving the overall situational awareness~\cite{INSperformance}.
During navigation the radar plays a key role in the formation of the crew's situational awareness and thus in dealing with ship encounter situations and in the decision-making for collision avoidance~\cite{Wu2021}. Through the Automatic Radar Plotting Aid (ARPA)~\cite{arpaPerformance95} the radar can automatically detect and calculate other ships' trajectories.
Integration between the radar system and the INS components is supported by a navigation network and by leveraging two standard network protocols: NMEA 0183~\cite{nmea0183} and ASTERIX CAT-240~\cite{asterix240}.
The former enables the interaction among all devices, the latter supports video data transmission between the radar antennas and the displays.

While these technologies contribute to improving the safety and effectiveness of navigation, they also expose ships to the Cybersecurity threat.
Meland et al.~\cite{Meland2021} presented an overview of 46 maritime cyber security incidents occurred in the last decade (2010-2020). 
While the overall number of cyber attacks is relatively small when compared to other sectors, unfortunately the impact of cyber attacks in the maritime sector can be very high.
The incident of the \emph{Ever Given}\footnote{\url{https://en.wikipedia.org/wiki/2021_Suez_Canal_obstruction}} in the Suez Canal, although not due to a cyber attack, is a dire reminder of the magnitude of the disruptions that can occur in the maritime sector.

An attacker can have a variety of objectives ranging from the ``mere'' disruption of the operations aiming to inflict hefty economic losses or the payment of a ransom to the deliberate attempt to cause a collision.
Since the crews make decisions by cross-checking between multiple systems and what they perceive when they look outside, an attack is likely to succeed when it takes this matter into account.

A key problem is that both the NMEA and ASTERIX protocols assume that the navigation network as well as the interconnected subsystems are trusted and no provision for cryptographic protection is therefore provided. The aforementioned cybersecurity incidents show that this trust assumption is no longer tenable and this security weakness of NMEA associated with the INS is widely recognized~\cite{INSIntegrity}. Even worse, due to the average life span of modern ships (up to 40 years) and the fact that retrofitting the INS is expensive and time-consuming, these weaknesses are here to stay. (It is however reasonable to expect that intrusion detection tools capable to identify unexpected network traffic and/or resource consumption deviations will eventually find their way into the INS.)

Yet, launching a successful cyber attack against a ship is not easy.
INSs are typically offline and penetrating them through lateral movements from other networks and controlling an attack from the Internet may not be an option.
Additionally, both the individual components and the configuration of the INS may vary from ship to ship.
For these reasons we argue that a cyber attack with reasonable chances of success requires the development of a malware that:
\begin{itemize}
\item exhibits a high degree of autonomy (e.g.~the ability to pursue its objectives without human support or guidance),
\item does not rely on the knowledge about the individual components and the actual configuration of the INS,
\item is stealthy, i.e.~its behavior is hard to detect by anomaly detection tools available at the host and/or network level (this implies that the malicious activity must be executed by requiring a moderate use of the CPU, of the memory resources and of network bandwidth to avoid behavioral fingerprinting) and---for the most sophisticated types of attacks---its effects are difficult to detect by the crew (this implies that the information shown on the displays is consistent with the other sources of information contributing to the situational awareness.)
\end{itemize}
Even if ships are offline during navigation, the injection of the malware into their INS can be carried out during management or upgrade of the INS.

Previous works partially consider the above constraints.
In general, they do not discuss the security of the ASTERIX protocols along with the internals about attacks against radar systems.

Hareide et al.~\cite{hareide2018enhancing} consider INSs as isolated systems.
They achieve a successful attack by using a USB key to inject their malware into the Windows workstation running the electronic chart system.
The malware can run without any external control, and they programmed it to trigger at a specific GPS place.
It leverages a man-in-the-middle (MITM) attack~\cite{MITREAttackMITM} to inject false GPS values and force the chart system to show a faulty position.

Casanovas et al.~\cite{casanovas2015vulnerability} analyzed an equivalent standard protocol for surveillance information exchange among different aerial traffic control centers, namely ASTERIX CAT-032.
They show how the lack of security mechanisms in such a protocol can lead to a MITM attack enabling the deletion and insertion of aircrafts and the update of their track, thus causing a misperception to air traffic control operators.

Kessler~\cite{kessler2019cybersecurity} reported that in late 2017 a cyber-consulting company successfully attacked a ship's radar.
After attacking the INS network from the Internet, they gained access to the radar workstation, altered the display by deleting targets, and thus blinding the ship. 

In this work, we introduce a novel class of attacks against maritime radar systems and we propose a method to detect them. 

First, the attacks can be performed by malware acting on its own, without command-and-control servers, and able to determine when the ship's state is suitable to execute them.
The malware can be easily adapted to each INS configuration.
Moreover, the malware only exploit security weaknesses and specific features of ASTERIX and NMEA protocols and the configuration of the INS network. 
They do not require access to the radar workstation.
The attacks can either corrupt and make the radar display unavailable or be sophisticated and stealthy up to modifying explicit details of the radar image in real-time and with extreme realism.
After modifying radar images, they generate consistent data for the other INS equipment.
Finally, we show the malware performs all the operations requiring very little CPU and memory resources and a limited network bandwidth.

The contributions of our work are as follows.
\begin{enumerate}
 \item We provide a high-level yet precise reconstruction of an Integrated Navigation System and its security assumptions.
 \item We argue that these assumptions are no longer justified in the light of emerging cyber threats and actors and impact at a successful attack.
 \item We show that security weaknesses can be exploited in such a way to disrupt situational awareness and lead to dramatic consequences.
 \item We argue that crafting an attack of this type requires a sophisticated and determined attacker, but given the severity of the impact (e.g., life loss, environment, or economic), the threat should not be underestimated.
 State actors and criminal organizations have already been shown to have the skills, resources, and determination to plan and execute attacks with this (and ever greater) level of sophistication.
 \item We present a network monitoring technique that detects such and unknown attacks against the radar system.
 It runs without requiring any changes to the existing INS configuration.
\end{enumerate}

This paper is structured as follows.
In Section~\ref{sec:background} we recall some preliminary notions.
In Section~\ref{sec:attack-techniques} we introduce the threat model and attack techniques to hijack a radar system.
In Section~\ref{sec:radarh}, we describe novel attacks exploiting the above techniques and in Section~\ref{sec:detection} a system to detect them.
In Section~\ref{sec:expeval}, we demonstrate the feasibility of the attacks and evaluate our detection system.
Finally, we conclude the paper in Section~\ref{sec:conclusion}.

%% file: sections/2-background/section.tex
\section{Background}
\label{sec:background}

In this section, we recall the notions that are relevant for correctly understanding the content of the paper.

\input{sections/2-background/network}
\input{sections/2-background/nmea}

\input{sections/2-background/ais}
\input{sections/2-background/radar_system}

\input{sections/2-background/asterix}

\input{sections/2-background/arpa}

\input{sections/2-background/colreg}

%% file: sections/2-background/network.tex
\subsection{Ship navigation network}
\label{sec:ship-navigation-network}
\begin{figure}[h]
    \centering
    \includegraphics[width=.5\textwidth]{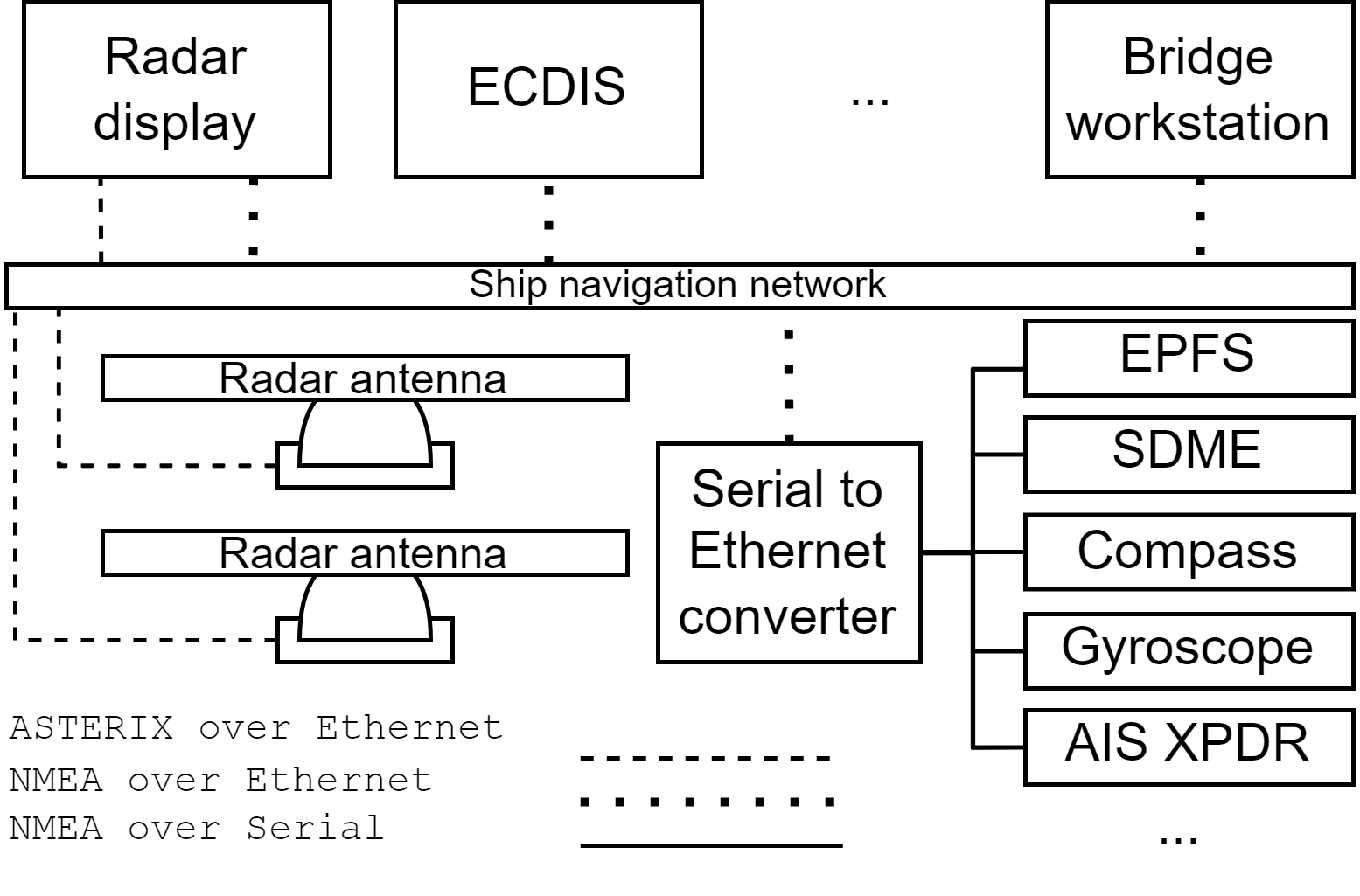}
    \caption{Ship navigation network topology.}
    \label{fig:network-topology}
\end{figure}

On a ship, the navigation network (see Figure \ref{fig:network-topology}) connects sensors and bridge systems.
Its typical configuration follows a homogeneous integration pattern in which multiple devices receive, process, and visualize data exchanged in a shared Ethernet network~\cite{INSIntegrity}, where any connected endpoint can listen and add its own messages to all broadcasted traffic. Similarly, any device can discover, listen and communicate with multicast flows via the standard IGMP protocol~\cite{IGMP}.

The main aim is to ease creating a system, namely an Integrated Navigation System (INS), that promotes data fusion and synergy between different equipment operating independently.

A Serial to Ethernet converter is a collection unit for sensors devices installed on a ship that forwards data to the navigation network. 
The main sensors devices are the Electronic Position Fixing System (EPFS), the Speed and Distance Measurement Equipment (SDME), the Compass, the Gyroscope, and the Automatic Identification System (AIS) transponder (see Section~\ref{sec:background-ais}).
The navigation network also hosts the two most essential navigational equipment: the radar system (see Section~\ref{sec:radarh-overview}) and a specialized digital navigation computer, namely the Electronic Chart Display and Information System (ECDIS).
Finally, one or more workstations are at the disposal of the deck personnel. 

NMEA 0183 and ASTERIX are the core standards that enable the integration of all these sensors and equipment.
We will briefly introduce them in the following sections.

%% file: sections/2-background/nmea.tex
\subsection{NMEA 0183}
\label{sec:background-nmea}

The NMEA 0183 standard~\cite{nmea0183} defines an electrical and data exchange format between maritime electronics.
Most of the sensor devices and systems installed on a ship communicate via NMEA~\cite{INSperformance}. 

Each message (or \textit{sentence}) is comprised of a start character followed by comma-delimited fields and a simple checksum terminated by a two-byte delimiter. 
Of particular interest is the \emph{talker sentence} format in which each message contains a two letter talker identifier, a three letter sentence type and a variable number of fields.
An example sentence with the talker identifier follows.
\begin{center}
    \texttt{\$HETHS,33.2,A*1F}
\end{center}
It represents a message emitted by the gyroscope (\texttt{HE}), with a sentence type related to the true heading and status (\texttt{THS}), and indicating a sensor heading measurement of $33.2^\circ$, sent automatically (\texttt{A}) and having a checksum of $\texttt{1F}_{16}=31_{10}$.

From a security standpoint, this protocol has no built-in message authentication nor any confidentiality protection.

%% file: sections/2-background/ais.tex
\subsection{AIS}
\label{sec:background-ais}

The Automatic Identification System (AIS)~\cite{ais} is a standard system for data exchange between ships and other maritime authorities.
Multiple standard message types are defined, covering a broad range of safety-enhancing functionality. For instance, periodically broadcasted position reports indicate the current course and speed of ships, reducing the risk of collisions.

Reception and transmission are carried out over Very High Frequency (VHF) radio data links. 
Each AIS message is usually transported within NMEA talker sentences with types \texttt{VDM} and \texttt{VDO}, respectively for received and sent messages. Such encapsulation is often used for displaying and utilizing the received information on other INS devices. In particular, radar plotters can associate their tracks with received AIS information~\cite{radarPerformance04}.

Since 2002, all ships engaged in international voyages above 300 gross tonnage~\cite{solas2020} are mandated to install an AIS, with other regulations suggesting its performance requirements~\cite{AISPerformance}, how to perform the mandatory annual test~\cite{AISTesting} and guidelines for its correct operation~\cite{AISGuidelines}.

The protocol presents no message authentication and - due to its broadcast nature - does not provide any mean to confidentially exchange information; albeit this issue has been investigated multiple times in the literature ~\cite{AISAuthenticated,AISAuthenticated2,AISProtected}, the few regulations concerning the hardening of AIS are only devoted to the protection of its radio frequency band from rogue transmitters, but do not provide any solution to secure the AIS from insecure inputs~\cite{AISIMOProtection}.

%% file: sections/2-background/radar_system.tex
\subsection{RADAR system}
\label{sec:radarsys}
RAdio Detection And Ranging (RADAR) is a system that can detect surrounding objects using radio waves. 
The whole radar system relies on different devices but we can look at it as composed by two main ones: an \emph{antenna} unit and a display unit, namely the \emph{Plan Position Indicator} (PPI).

An antenna rotates 360 degrees about its vertical axis, radiating waves and receiving returning echoes from targets.
Antennas have their own specifications that differ between manufacturers.
In particular, each specification includes the \emph{rotation speed}, and a resolution related to the \emph{bearing} and \emph{range}.
The rotation speed specifies the speed at which an antenna is rotated by the motor.
The bearing resolution, or \emph{angular} resolution, determines the ability of a radar system to separate targets at the same distance but at different direction.
The range resolution determines the ability to resolve between two targets on the same direction, but at slightly different distances.

Echoes from an antenna can be transmitted to the PPI via Ethernet network using proprietary solutions~\cite{radarBicycle,radarPI} or the standard protocol from ASTERIX (see Section~\ref{sec:asterix}).

The PPI is a circular display representing the antenna, with the own ship in the center.
A radial trace sweeps in unison with the radar antenna around the central point.
Each trace represents echo signals in plan position with bearing and range displayed in polar coordinates.
The top of the display may be configured to represent different perspectives. 
In the \emph{head-up} mode, the zero of the PPI represents the own ship's course, and the bearing of the displayed targets will be relative to its heading.
In the \emph{north-up} mode, the zero represents the true north, a heading marker represents the true course of the own ship, and all bearings of targets are actual.

Digital PPIs must emulate the behavior of traditional radar scopes.
In particular, every echo received must persist on the PPI for at least the time of half a rotation~\cite{radarManual}.
Moreover, standard regulations state that if a PPI receives multiple traces for the same rotation angle during the persistence time interval, it has to sum their echoes~\cite[\S15.6.3.2.e]{radarTesting}.

Digital PPIs also add new capabilities over traditional radar scopes.
For example, the echo \emph{trail} is used to visually understand the movements of other vessels, i.e., path and speed, by displaying a residual image at different times of an echo.
Radar systems that support ARPA capabilities (see Section~\ref{sec:background-arpa}) can automatically provide an accurate estimate of such movements.

%% file: sections/2-background/asterix.tex
\subsection{ASTERIX}
\label{sec:asterix}

ASTERIX~\cite{asterix} is a suite of standard protocols for data exchange of radar information between systems proposed by EUROCONTROL.
The ASTERIX standards identify a collection of message types, called categories or CAT.
Of particular interest for this work is CAT-240, i.e., Radar Video Transmission~\cite{asterix240}, used to transfer video data from antennas to Plan Position Indicator (PPI) displays.
After its specification in 2009, ASTERIX CAT-240 has been adopted by manufacturers as the de-facto network video standard~\cite{johnson2013using}.

As sketched in Figure~\ref{fig:asterix-protocol}, each CAT-240 message combines a header and a video block and is related to an angle span. 
The header provides information about the block and metadata like time of day or the \emph{System Identification Code} and \emph{System Area Code} (SIC/SAC) that identify the transmitting antenna.
Once decoded, the video block is a sequence of cells located on a polar coordinate system centered around the position of the transmitting antenna.
The angle span is between \emph{start\_az} and \emph{end\_az}.
Cells indicate the echo strength quantized using cell\_res bits.
Moreover, each cell starts at a distance $\rho$ that can be calculated by leveraging their homogeneity among the distance direction as
$ \rho = D * (b + i) * c / 2$ where $D$ and $b$ are included in the header and represent the \emph{cell duration} parameter and the \emph{center bias}, respectively, while $i$ is the cell index (0-based), and $c$ is the light celerity\footnote{$299792458 \; m/s$ as in~\cite{asterix240}}.

Referring to the resolution of antennas (see Section~\ref{sec:radarh-overview}), the bearing resolution determines the minimum span between $start\_az$ and $end\_az$ while the range resolution determines the minimum $cell\_dur$.

Finally, $message\_id$ is a sequence number used by the receiver to reorder packets.

From a security standpoint, as emphasized in~\cite{janvcik2019security,ASTERIXAttack}, the ASTERIX protocol does not implement any authentication and encryption features.
\input{sections/2-background/asterix_drawing}

%% file: sections/2-background/asterix_drawing.tex
\tikzmath{
    \begAzimuth = 20;
    \endAzimuth = 60;
    \indexAzimuth = \begAzimuth - 10;
    \midAzimuth = (\endAzimuth + \begAzimuth) / 2;
    \circleRadius = 3;
    \cellRadius = \circleRadius / 4;
    \cellOneBegRadius = 0.0;
    \cellOneEndRadius = \cellOneBegRadius + \cellRadius;
    \cellOneIBegX = \cellOneBegRadius * sin(\endAzimuth);
    \cellOneIBegY = \cellOneBegRadius * cos(\endAzimuth);
    \cellOneIMidX = \cellOneBegRadius * sin(\midAzimuth);
    \cellOneIMidY = \cellOneBegRadius * cos(\midAzimuth);
    \cellOneIEndX = \cellOneBegRadius * sin(\begAzimuth);
    \cellOneIEndY = \cellOneBegRadius * cos(\begAzimuth);
    \cellOneOBegX = \cellOneEndRadius * sin(\begAzimuth);
    \cellOneOBegY = \cellOneEndRadius * cos(\begAzimuth);
    \cellOneOMidX = \cellOneEndRadius * sin(\midAzimuth);
    \cellOneOMidY = \cellOneEndRadius * cos(\midAzimuth);
    \cellOneOEndX = \cellOneEndRadius * sin(\endAzimuth);
    \cellOneOEndY = \cellOneEndRadius * cos(\endAzimuth);
    \cellTwoBegRadius = \cellOneEndRadius;
    \cellTwoEndRadius = \cellTwoBegRadius + \cellRadius;
    \cellTwoIBegX = \cellTwoBegRadius * sin(\endAzimuth);
    \cellTwoIBegY = \cellTwoBegRadius * cos(\endAzimuth);
    \cellTwoIMidX = \cellTwoBegRadius * sin(\midAzimuth);
    \cellTwoIMidY = \cellTwoBegRadius * cos(\midAzimuth);
    \cellTwoIEndX = \cellTwoBegRadius * sin(\begAzimuth);
    \cellTwoIEndY = \cellTwoBegRadius * cos(\begAzimuth);
    \cellTwoOBegX = \cellTwoEndRadius * sin(\begAzimuth);
    \cellTwoOBegY = \cellTwoEndRadius * cos(\begAzimuth);
    \cellTwoOMidX = \cellTwoEndRadius * sin(\midAzimuth);
    \cellTwoOMidY = \cellTwoEndRadius * cos(\midAzimuth);
    \cellTwoOEndX = \cellTwoEndRadius * sin(\endAzimuth);
    \cellTwoOEndY = \cellTwoEndRadius * cos(\endAzimuth);
    \cellTwoIndexX = (\cellTwoBegRadius + \cellTwoEndRadius) / 2 * sin(\indexAzimuth);
    \cellTwoIndexY = (\cellTwoBegRadius + \cellTwoEndRadius) / 2 * cos(\indexAzimuth);
    \cellThreeBegRadius = \cellTwoEndRadius;
    \cellThreeEndRadius = \cellThreeBegRadius + \cellRadius;
    \cellThreeIBegX = \cellThreeBegRadius * sin(\endAzimuth);
    \cellThreeIBegY = \cellThreeBegRadius * cos(\endAzimuth);
    \cellThreeIMidX = \cellThreeBegRadius * sin(\midAzimuth);
    \cellThreeIMidY = \cellThreeBegRadius * cos(\midAzimuth);
    \cellThreeIEndX = \cellThreeBegRadius * sin(\begAzimuth);
    \cellThreeIEndY = \cellThreeBegRadius * cos(\begAzimuth);
    \cellThreeOBegX = \cellThreeEndRadius * sin(\begAzimuth);
    \cellThreeOBegY = \cellThreeEndRadius * cos(\begAzimuth);
    \cellThreeOMidX = \cellThreeEndRadius * sin(\midAzimuth);
    \cellThreeOMidY = \cellThreeEndRadius * cos(\midAzimuth);
    \cellThreeOEndX = \cellThreeEndRadius * sin(\endAzimuth);
    \cellThreeOEndY = \cellThreeEndRadius * cos(\endAzimuth);
    \cellThreeIndexX = (\cellThreeBegRadius + \cellThreeEndRadius) / 2 * sin(\indexAzimuth);
    \cellThreeIndexY = (\cellThreeBegRadius + \cellThreeEndRadius) / 2 * cos(\indexAzimuth);
    \cellFourBegRadius = \cellThreeEndRadius;
    \cellFourEndRadius = \cellFourBegRadius + \cellRadius;
    \cellFourIBegX = \cellFourBegRadius * sin(\endAzimuth);
    \cellFourIBegY = \cellFourBegRadius * cos(\endAzimuth);
    \cellFourIMidX = \cellFourBegRadius * sin(\midAzimuth);
    \cellFourIMidY = \cellFourBegRadius * cos(\midAzimuth);
    \cellFourIEndX = \cellFourBegRadius * sin(\begAzimuth);
    \cellFourIEndY = \cellFourBegRadius * cos(\begAzimuth);
    \cellFourOBegX = \cellFourEndRadius * sin(\begAzimuth);
    \cellFourOBegY = \cellFourEndRadius * cos(\begAzimuth);
    \cellFourOMidX = \cellFourEndRadius * sin(\midAzimuth);
    \cellFourOMidY = \cellFourEndRadius * cos(\midAzimuth);
    \cellFourOEndX = \cellFourEndRadius * sin(\endAzimuth);
    \cellFourOEndY = \cellFourEndRadius * cos(\endAzimuth);
    \cellFourIndexX = (\cellFourBegRadius + \cellFourEndRadius) / 2 * sin(\indexAzimuth);
    \cellFourIndexY = (\cellFourBegRadius + \cellFourEndRadius) / 2 * cos(\indexAzimuth);
    \azSpanRadius = \circleRadius * 1.1;
    \azSpanBegX = \azSpanRadius * sin(\begAzimuth);
    \azSpanBegY = \azSpanRadius * cos(\begAzimuth);
    \azSpanMidX = \azSpanRadius * sin(\midAzimuth);
    \azSpanMidY = \azSpanRadius * cos(\midAzimuth);
    \azSpanEndX = \azSpanRadius * sin(\endAzimuth);
    \azSpanEndY = \azSpanRadius * cos(\endAzimuth);
}
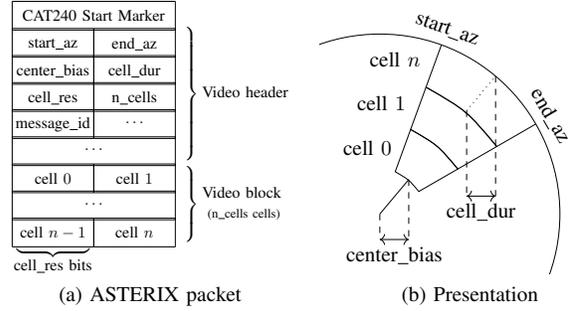
\begin{figure}
    \centering
    \subfloat[ASTERIX packet]{
        \resizebox{!}{0.20\textwidth}{
        \begin{bytefield}[bitformatting={},bitwidth=auto]{8}
        \bitbox{8}{CAT240 Start Marker} \\
        \begin{rightwordgroup}{Video header}
            \bitbox{4}{start\_az}\bitbox{4}{end\_az} \\
            \bitbox{4}{center\_bias}\bitbox{4}{cell\_dur} \\
            \bitbox{4}{cell\_res}\bitbox{4}{n\_cells} \\
            \bitbox{4}{message\_id}\bitbox{4}{$\ldots$} \\
            \bitbox{8}{$\ldots$}
        \end{rightwordgroup} \\
        \begin{rightwordgroup}{Video block \\ { \footnotesize (n\_cells cells) }}
            \bitbox{4}{cell $0$}\bitbox{4}{cell $1$} \\
            \bitbox{8}{$\ldots$} \\
            \bitbox{4}{cell $n-1$}\bitbox{4}{cell $n$}
        \end{rightwordgroup} \\
        \bitbox[]{4}{$\underbrace{\hspace{1.6cm}}_{\text{\normalsize cell\_res bits}}$}
        \end{bytefield}
        }
    }
    \subfloat[\centering Presentation]{
        \resizebox{!}{0.22\textwidth}{
        \begin{tikzpicture}
        \clip (-1,-1) rectangle (4,4);
        \draw (0,0) circle (\circleRadius);
        \draw 
            (\cellTwoIBegX,\cellTwoIBegY) .. controls (\cellTwoIMidX, \cellTwoIMidY) .. (\cellTwoIEndX,\cellTwoIEndY) --
            (\cellTwoOBegX,\cellTwoOBegY) .. controls (\cellTwoOMidX, \cellTwoOMidY) .. (\cellTwoOEndX,\cellTwoOEndY) -- (\cellTwoIBegX,\cellTwoIBegY);
        \draw 
            (\cellThreeIBegX,\cellThreeIBegY) .. controls (\cellThreeIMidX, \cellThreeIMidY) .. (\cellThreeIEndX,\cellThreeIEndY) --
            (\cellThreeOBegX,\cellThreeOBegY) .. controls (\cellThreeOMidX, \cellThreeOMidY) .. (\cellThreeOEndX,\cellThreeOEndY) -- (\cellThreeIBegX,\cellThreeIBegY);
        \draw 
            (\cellFourIBegX,\cellFourIBegY) .. controls (\cellFourIMidX, \cellFourIMidY) .. (\cellFourIEndX,\cellFourIEndY);
        \draw (\cellFourIEndX,\cellFourIEndY) -- (\cellFourOBegX,\cellFourOBegY);
        \draw (\cellFourOEndX,\cellFourOEndY) -- (\cellFourIBegX,\cellFourIBegY);
        \draw (\azSpanBegX,\azSpanBegY) node [rotate=360-\begAzimuth] {start\_az};
        \draw (\azSpanEndX,\azSpanEndY) node [rotate=360-\endAzimuth] {end\_az};
        \draw[-] (\cellOneIMidX,\cellOneIMidY) -- (\cellOneOMidX, \cellOneOMidY);
        \draw[dashed] (\cellOneIMidX,\cellOneIMidY) -- (\cellOneIMidX, \cellOneOMidY-1);
        \draw[dashed] (\cellOneOMidX,\cellOneOMidY) -- (\cellOneOMidX, \cellOneOMidY-1);
        \draw[<->] (\cellOneIMidX, \cellOneOMidY-1) -- node [below] {center\_bias} (\cellOneOMidX, \cellOneOMidY-1);
        \draw (\cellTwoIndexX, \cellTwoIndexY) node [rotate=0,xshift=-0.4cm] {cell $0$};
        \draw (\cellThreeIndexX, \cellThreeIndexY) node [rotate=0,xshift=-0.3cm] {cell $1$};
        \draw (\cellFourIndexX, \cellFourIndexY) node [rotate=0,xshift=-0.2cm] {cell $n$};
        \draw[dotted] (\cellFourIMidX,\cellFourIMidY) -- (\cellFourOMidX, \cellFourOMidY);
        \draw[dashed] (\cellFourIMidX,\cellFourIMidY) -- (\cellFourIMidX, \cellFourOMidY-2);
        \draw[dashed] (\cellFourOMidX,\cellFourOMidY) -- (\cellFourOMidX, \cellFourOMidY-2);
        \draw[<->] (\cellFourIMidX, \cellFourOMidY-2) -- node [below] {cell\_dur} (\cellFourOMidX, \cellFourOMidY-2);
        \end{tikzpicture}
        }
    }
    \caption{Correspondence between ASTERIX and the PPI.}
    \label{fig:asterix-protocol}
\end{figure}

%% file: sections/2-background/arpa.tex
\subsection{ARPA}
\label{sec:background-arpa}
The Automatic Radar Plotting Aid (ARPA) is a system integrated with radar displays that carry out the task of \textit{radar plotting}.
International law mandates ships exceeding 10000 gross tonnage~\cite[V\S2.8]{solas2020} to be equipped with such a device.

Radar plotting allows a radar officer to follow a target over time, reconstructing its trajectory w.r.t. the own ship, estimating its course, speed, range at the closest point of approach (DCPA), and the predicted time to CPA (TCPA). 
It is worth noting that when TCPA is a negative number, it signals an increasing trend, i.e. the target CPA is getting further from the ship.

A target can be acquired manually by an officer or automatically when it enters in configurable \emph{acquisition zones}.
A target becomes acquired after it persists for 5 out of 10 consecutive scans~\cite[\S3.3.3]{radarPerformance04}.

\input{sections/2-background/arpa_drawing}
 
Once acquisition is performed, the radar system tries to follow the movement of targets inside of the image. This process, combined with the EPFS and SDME data allows the ARPA system to estimate a target's trajectory. Finally, if estimation is successful a target becomes a \textit{tracked target}~\cite[Ann.G]{radarTesting}. 
Targets that are being tracked appear on display with the symbol depicted in Figure~\ref{fig:arpa-targets}a.
Within INS, the radar system propagates information about such targets via NMEA and using the \texttt{TTM} (Tracked Target Message) sentences.

ARPA constantly evaluates the CPA and the TCPA status of each tracked target.
Acquired targets that move inside the \emph{guard zone}, i.e., a zone configured with a given radius (CPA) and time threshold (TCPA), generate an alarm.
Targets that generate an alarm appear on display as dangerous and with the symbol depicted in Figure~\ref{fig:arpa-targets}b.

Finally, a tracked target is judged as a lost target when no return is received for nine consecutive scans and appears on display with the symbol depicted in Figure~\ref{fig:arpa-targets}c.

%% file: sections/2-background/arpa_drawing.tex
\begin{figure}[h]
    \centering
    \subfloat[\centering Acquired]{
        \resizebox{!}{0.10\textwidth}{
        \definecolor{darkgreen}{rgb}{0.0, 0.7, 0.03}
        \begin{tikzpicture}
            \draw[darkgreen,line width=3pt] (0,0) circle (1);
            \draw[darkgreen,line width=3pt] (0,0) -- (0.707,0.707);
            \draw[darkgreen,dash pattern=on 8pt off 16pt, line width=3pt] (0.707,0.707) -- (4,4);
        \end{tikzpicture}
        }
    }
    \subfloat[\centering Dangerous]{
        \resizebox{!}{0.10\textwidth}{
        \definecolor{dangerred}{rgb}{0.9, 0.0, 0.0}
        \begin{tikzpicture}
            \draw[dangerred,line width=5pt] (0,0) circle (1);
            \draw[dangerred,line width=3pt] (0,0) -- (0.707,0.707);
            \draw[dangerred,dash pattern=on 8pt off 16pt, line width=3pt] (0.707,0.707) -- (4,4);
        \end{tikzpicture}
        }
    }
    \subfloat[\centering Lost]{
        \resizebox{!}{0.04\textwidth}{
        \definecolor{darkgreen}{rgb}{0.0, 0.7, 0.03}
        \definecolor{dangerred}{rgb}{0.9, 0.0, 0.0}
        \begin{tikzpicture}
            \draw[darkgreen,line width=3pt] (0,0) circle (1);
            \draw[darkgreen,line width=3pt] (0,0) -- (0.707,0.707);
            \draw[dangerred,line width=3pt] (-1,-1) -- (1,1);
            \draw[dangerred,line width=3pt] (-1,1) -- (1,-1);
        \end{tikzpicture}
        }
    }
    \caption{ARPA target symbols on the PPI.}
    \label{fig:arpa-targets}
\end{figure}
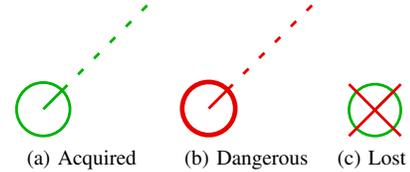

%% file: sections/2-background/colreg.tex
\subsection{COLREGs}
\label{sec:colregs}

In maritime navigation, vessels should obey the International Regulations for Preventing Collisions at Sea, namely COLlision REGulations (COLREGs), agreed to by the IMO in 1972~\cite{colregs}.
These rules specify maneuvers that ships must take in situations where a risk of collision occurs.
A vessel may employ radar and ARPA to assess its relative position, angle of approach, and speed against near ships and determine such a risk. 

This work addresses two COLREGs rules where only one of the involved vessels must maneuver: \emph{overtaking}, and \emph{crossing} situations.

\input{sections/2-background/colreg_drawing}

In particular, rule 13 is about overtaking and states that "a vessel shall be deemed to be overtaking when coming up with another vessel from a direction more than 22.5 degrees abaft her beam".
Fig.~\ref{fig:colregs}a illustrates a overtaking situation.
In such a situation, the vessel $a$ must overtake $b$, and common practice on the water dictates that the overtaking boat should pass on the right-hand side of the slower vessel $b$.

Rule 15 governs crossing situations and states that “when two power-driven vessels are crossing so as to involve risk of collision, the vessel which has the other on her own starboard side shall keep out of the way and shall, if the circumstances of the case admit, avoid crossing ahead of the other vessel”.
Fig.~\ref{fig:colregs}b illustrates a crossing situation.
In such a situation, the vessel $a$ is in a collision course with $b$ and must veer to its starboard so that it does not cross ahead of $b$.

%% file: sections/2-background/colreg_drawing.tex
\begin{figure}[h]
    \centering
    \subfloat[Overtaking]{%
        \resizebox{0.22\textwidth}{!}{
        \begin{tikzpicture}
            \tikzstyle{every node}=[font=\fontsize{70}{0}\selectfont]
            \begin{scope}[rotate around={270:(0,0)}]
            \fill[lightgray]
                ({ -6 * cos(-22.5) }, -4) -- ({ -6 * cos(-22.5) }, { 10 + 5 * sin(-22.5) }) -- (-1, 10) -- (1, 10) -- ({ 6 * cos(-22.5) }, { 10 + 5 * sin(-22.5) }) -- ({ 6 * cos(-22.5) }, -4);
            \draw[fill=white] (-1,-3) rectangle (1, 3) node[pos=.5] {a};
            \draw (-1, 10) -- ({ -6 * cos(-22.5) }, { 10 + 5 * sin(-22.5) });
            \draw ( 1, 10) -- ({ -6 * cos(180.0+22.5) }, { 10 + 5 * sin(180.0+22.5) });
            \draw[dashed] (6,10) -- (-6, 10);
            \draw[-{Latex[scale=5]}]
                (-6, 10)
                    .. controls ({ -6 * cos(-11.25) }, { 10 + 5 * sin(-11.25) }) ..
                    node [midway, above, scale=2, yshift=0.2cm] {$22.5^\circ$}
                ({ -6 * cos(-22.5) }, { 10 + 5 * sin(-22.5) });
            \draw[-{Latex[scale=5]}]
                (6, 10)
                    .. controls ({ 6 * cos(-11.25) }, { 10 + 5 * sin(-11.25) }) ..
                    node [midway, below, scale=2, yshift=-0.2cm] {$22.5^\circ$}
                ({ 6 * cos(-22.5) }, { 10 + 5 * sin(-22.5) });
            \draw[fill=white]  (-1,10) rectangle (1, 16) node[pos=.5] {b};
            \draw[dashed, ultra thick] (0, 16) -- (0, 19);
            \draw[ultra thick, red, -{Latex[scale=2]}, dashed] (-0.25, 4) .. controls (-0.25, 6) .. (-3, 10) .. controls (-4, 12) .. (-4, 18);
            \draw[ultra thick, red, -{Latex[scale=2]}] ( 0.25, 4) .. controls ( 0.25, 6) .. ( 3, 10) .. controls ( 4, 12) .. ( 4, 18);
            \end{scope}
        \end{tikzpicture}
        }
    }
    \subfloat[Crossing]{%
        \resizebox{!}{0.125\textwidth}{
        \begin{tikzpicture}
            \tikzstyle{every node}=[font=\fontsize{18}{0}\selectfont]
            
            \draw (0,-0.25) rectangle (1.25, 0.25) node[pos=.5] {a};
            \draw[dotted] (1.25, 0) -- (5.0, 0);
            \draw[->,red] (1.40, -0.15) .. controls (3, -0.15) .. (6, -2.6);
            
            \begin{scope}[rotate around={-35:(5.5,-1.5)}]
                \draw (5,-1.25) rectangle (6.25, -1.75) node[sloped,pos=.5] {b};
                \draw[dotted] (5,-1.5) -- (2,-1.5);
            \end{scope}
        \end{tikzpicture}
        }
    }
    \caption{Illustration of COLREGs situations.}
    \label{fig:colregs}
\end{figure}
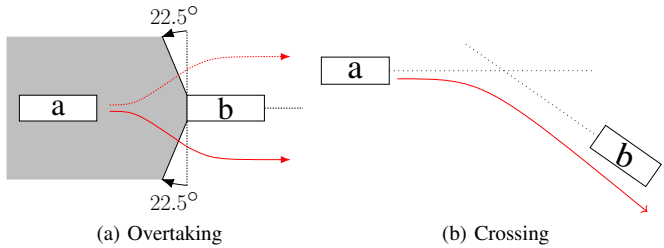

%% file: sections/3-techniques/section.tex
\section{Attack Techniques}
\label{sec:attack-techniques}

In this section, we present the threat model and the techniques to hijack a radar system.

\subsection{Overview}
In Sections~\ref{sec:asterix} and ~\ref{sec:background-nmea}, we highlight that both ASTERIX and NMEA protocols do not support confidentiality and authentication between communicating components.
Moreover, modern vessels are equipped with an INS (see Section~\ref{sec:ship-navigation-network}) where such components communicate through multicasts or broadcasts that allow anyone connected to listen to the exchanged packets.
The feature of these protocols and INS configuration represent the attack vector.

A coarse-grained attack can merely inject ASTERIX packets with false echoes to hide or corrupt the image displayed on the PPI.
This attack requires a little effort, generates the malfunction of an essential system for navigation, and yields a significant impact.
In particular, it can pose moderate risks to the ship's operations and safety and could force the start of emergency procedures to return the vessel to port.

Nevertheless, we here consider a novel kind of adversary capable of executing \textsl{fine-grained} attacks.
A fine-grained attack does not create malfunctions but alters the radar information without being detected.
It actively monitors the ship's status and only activates when potentially dangerous situations can happen.
After it activates, it can operate in real-time on specific areas of the radar image, and the changes appear realistic to the operators.
During the attack, it generates an amount of network traffic that does not appear anomalous compared to the one generated during the regular operation of the radar system. 
Moreover, it must perform all the above operations leveraging resources of INS components that may be limited in computing resources and with different hardware and operating systems.
This attack relies on a deep and specific knowledge of the domain it operates in and can pose a severe or catastrophic adverse effect, e.g., harm to individuals, major damage to the vessel and environment, and major financial loss. 

In the sections below, we present the assumptions under which adversaries operate and their techniques to perform from coarse-grained to fine-grained attacks.

\subsection{Threat model}
In this work, we focus on stealthy malware that accesses the ship navigation network to perform malicious activities.
The attacker's goal is to reduce the \textsl{situational awareness} of the ship officers to cause a disruption in operation or to significantly increase the probability of a safety-critical incident.

We assume that the ship under attack is equipped with an INS configured as detailed in Section~\ref{sec:ship-navigation-network}.
In particular, we assume it hosts a radar system compliant with regulations, performance standards, and behaviors as described in Section~\ref{sec:radarsys}.

Although some INS possess external connection capabilities~\cite{INSIntegrity}, we assume that the security of the navigation network is enforced with a restrictive policy, i.e., physically disconnected from other networks, including the Internet.

\paragraph*{Attacker's requirements}

We consider an adversary at least as a professional actor capable of gathering solid knowledge for generating or testing a novel attack.
The framework for Maritime Cyber-Risk Assessment (MACRA)~\cite[T.1]{macra} models such an adversary as a $\textit{Tier}_3$ attacker.
The abilities and resources of a $Tier_3$ attacker allow the adversary to leverage the maintenance operations or the supply chain compromise technique~\cite[T1195]{MITREAttack} to install the malware.
As reported in~\cite{INSIntegrity}, multiple components of an INS are periodically updated, as well as bridge workstations present vulnerabilities that might allow the installation of malware~\cite{raisingAwarenessECDIS,INSAttack,ECDISAttack} during a regular maintenance operation~\cite[p.32-36]{GuidelinesOnCyberSecurity}.

Once installed, the malware must operate stealthily and under the assumption that the network is isolated from the Internet.
Traditional malware that drop additional malicious payloads and require a command and control server are out of scope.
Instead, the attack requires a targeted malware~\cite{CAPEC-542} that can operate in autonomy and takes advantage of the specific technology environment.

\paragraph*{Attacker's capabilities}
Under the above assumptions, the adversary has different capabilities as follows.
Since the malware runs on a host connected to the na\-vi\-ga\-tion network, it can overhear the cleartext NMEA and ASTERIX packets like any other INS component.
NMEA traffic allows the malware to reconstruct and update the
ship’s state under attack by monitoring sensor devices and ARPA data.
For example, it can monitor its position, bearing, speed, nearby vessels, or targets acquired by radar operators.
ASTERIX traffic allows the malware to know what the PPI is displaying.

Furthermore, the malware can impersonate legitimate sensor devices and radar antennas by leveraging the lack of any authentication in such protocols.
For example, the malware can inject NMEA packets holding sentences with fake values from the Compass or the AIS transponder.
The injected NMEA packets appear as legitimate data to NMEA devices.
Likewise, the malware can inject ASTERIX packets holding messages with fake echoes to hijack the radar system.
We discuss radar hijacking techniques in the section below.

\subsection{Radar hijacking techniques}
\label{sec:hijacking_tech}

An adversary executes a radar hijacking attack to obtain the capabilities to \emph{add} and \emph{delete} targets on the PPI.
Under the assumption that the PPI behavior follows standard regulations, we remind that it must satisfy two conditions (see Section~\ref{sec:radarh}): $(i)$ echoes must persist for at least the time of half a rotation, and $(ii)$ if it receives a packet that overrides echoes during their persistence time, the PPI must \emph{sum} old and new values. 

Radar hijacking attacks leverage the injection of fake ASTERIX packets to modify echoes \emph{immediately after} the PPI receives the actual values from the legitimate antenna.
As a result of the two conditions above, the PPI always sums the fake and actual values.
It is worth noting that the behavior from standard regulations restricts an attacker only to increase the strength of existing echoes, thereby only enabling the capabilities to \emph{add} targets.

We experimented with the above restriction on the commercial radar of our testbed. 
Our tests consist in trying to delete the radar image by injecting the original packets after we update them with zero strength echoes.
The results showed that the PPI complies with the standard regulations as it sums values and prevents deleting the radar image.

In Figure~\ref{fig:htech}a, we show an example of an attack that the malware can exploit using only the capability to add a target.
For the sake of simplicity, we consider ASTERIX packets carrying a video block of six cells and related to the minimum angle span constrained by the bearing resolution of the antenna (e.g., one degree).
We reduce the echo strengths to on/off values.
This attack aims to add a fake echo to a trace that the PPI visualizes at a specific azimuth $\alpha$.
We assume that the PPI receives at time $t_0$ the ASTERIX message $(1a)$ for the azimuth $\alpha$ from the legitimate antenna.
Consequently, the PPI visualizes a trace with the two echoes that the message holds in the third and fifth cells.
In the meantime, the attacker can overhear $(1a)$ and create a new ASTERIX message $(2a)$ containing original echoes and the fake one in the sixth cell.
Then, the attacker can inject the new message $(2a)$ into the navigation network at time $t_1 = t_0 + \epsilon$, where $\epsilon$ is a small delay due to the attacker's operations.
After the PPI receives $(2a)$, it visualizes the new trace $(3a)$ for $\alpha$ that sums the echoes of $(2a)$ with the ones of $(1a)$ cell by cell.
Since $\epsilon$ is a negligible lag, i.e., in the order of milliseconds, a radar operator will not perceive the update.

To obtain the capability to delete targets, an attacker must create an outlier situation that a PPI handles by violating standard regulations.
We obtained such a condition with the commercial radar of our testbed by applying a standard feature of the ASTERIX protocol.
We injected packets that differ from the originals in the value of the echo strengths and the number of cells they contain.
In particular, we decreased it by shifting the \emph{center\_bias} parameter by one in the packets header (see Section~\ref{sec:asterix}).
Due to this difference, the PPI under test replaces the displayed echoes with the most recent data of the injected packet, thus allowing us to acquire the capability of deleting existing echoes.

\input{sections/3-techniques/drawing}
In Figure~\ref{fig:htech}b, we show an example of an attack using the \emph{center\_bias} parameter.
The objective of this second attack is to replace the trace generated by $(1b)$ for $\alpha$ with a new one that deletes the echo of the third cell and adds an echo to the sixth cell.
To this aim, the attacker creates the message $(2b)$ containing the two cells with echoes and with a shift of four cells from the center, i.e., \text{center bias = 4}.
When the PPI receives such a message, it should keep values of the first four cells of the visualized track $(1b)$ and sum the last two values of $(1b)$ with the ones of $(2b)$.
Instead, the PPI replaces the existing trace with $(3b)$ that corresponds to the most recent message $(2b)$, thus deleting the echo in the third cell.

We stress that an adversary can perform an attack that adds a target against any radar system. 
In contrast, the feasibility of an attack that deletes a target depends on the vendor-specific implementation of the PPI when the outlier situation we introduced above occurs.
In Section~\ref{sec:recon}, we show that the malware can automatically infer if the PPI under attack suffers from behavior similar to our testbed and allows attackers the delete capability.

%% file: sections/3-techniques/drawing.tex
\tikzmath{
    \circleRadius = 3;
    \rectangleBegX = 0;
    \rectangleSize = 0.5;
    \rectangleOneBegX = \rectangleBegX;
    \rectangleTwoBegX = \rectangleOneBegX + \rectangleSize;
    \rectangleThreeBegX = \rectangleTwoBegX + \rectangleSize;
    \rectangleFourBegX = \rectangleThreeBegX + \rectangleSize;
    \rectangleFiveBegX = \rectangleFourBegX + \rectangleSize;
    \rectangleSixBegX = \rectangleFiveBegX + \rectangleSize;
    \rectangleSixEndX = \rectangleSixBegX + \rectangleSize;
    \attackY = 1.25;
    \originalY = 2.25;
    \resultY = 0;
    \resultAzimuthBeg = 70;
    \resultAzimuthEnd = 80;
    \resultRadius = \rectangleSize * 6;
    \cellOneRadius = 1 * (\resultRadius / 6);
    \cellTwoRadius = 2 * (\resultRadius / 6);
    \cellThreeRadius = 3 * (\resultRadius / 6);
    \cellFourRadius = 4 * (\resultRadius / 6);
    \cellFiveRadius = 5 * (\resultRadius / 6);
}
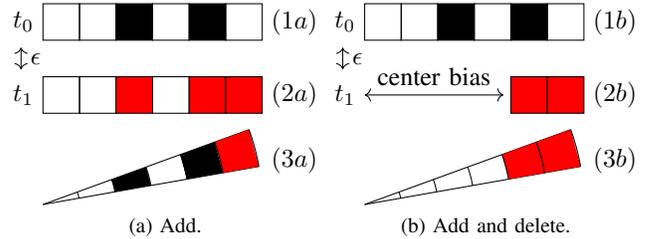
\begin{figure}[h]
    \centering
    \subfloat[Add.]{\resizebox{!}{0.15\textwidth}{\begin{tikzpicture}
    \draw (\rectangleOneBegX,\originalY) node[yshift=0.25cm,left] {$t_0$} rectangle +(\rectangleSize,\rectangleSize);
    \draw (\rectangleTwoBegX,\originalY) rectangle +(\rectangleSize,\rectangleSize);
    \draw[fill=black] (\rectangleThreeBegX,\originalY) rectangle +(\rectangleSize,\rectangleSize);
    \draw (\rectangleFourBegX,\originalY) rectangle +(\rectangleSize,\rectangleSize);
    \draw[fill=black] (\rectangleFiveBegX,\originalY) rectangle +(\rectangleSize,\rectangleSize);
    \draw (\rectangleSixBegX,\originalY) rectangle +(\rectangleSize,\rectangleSize);
    
    \draw (\rectangleSixEndX,\originalY) node[yshift=0.25cm,right] {$(1a)$};
    \draw (\rectangleOneBegX,\attackY) node[yshift=0.25cm,left] {$t_1$} rectangle +(\rectangleSize,\rectangleSize);
    \draw (\rectangleTwoBegX,\attackY) rectangle +(\rectangleSize,\rectangleSize);
    \draw[fill=red] (\rectangleThreeBegX,\attackY) rectangle +(\rectangleSize,\rectangleSize);
    \draw (\rectangleFourBegX,\attackY) rectangle +(\rectangleSize,\rectangleSize);
    \draw[fill=red] (\rectangleFiveBegX,\attackY) rectangle +(\rectangleSize,\rectangleSize);
    \draw[fill=red] (\rectangleSixBegX,\attackY) rectangle +(\rectangleSize,\rectangleSize);
    
    \draw (\rectangleSixEndX,\attackY) node[yshift=0.25cm,right] {$(2a)$};
    \draw[<->] ({\rectangleBegX - 0.3}, { \originalY - 0.1 }) -- node [xshift=0.2cm,centered] {$\epsilon$} ({\rectangleBegX - 0.3}, { \attackY + \rectangleSize + 0.1 });
    \draw ({ \rectangleBegX + 6 * \rectangleSize },\resultY) node[yshift=0.6cm,right] {$(3a)$};
    \draw (0,0) -- ({ \resultRadius*sin(\resultAzimuthBeg) }, { \resultRadius*cos(\resultAzimuthBeg) });
    \draw (0,0) -- ({ \resultRadius*sin(\resultAzimuthEnd) }, { \resultRadius*cos(\resultAzimuthEnd) });
    \path[fill=black] ({\cellTwoRadius*sin(\resultAzimuthBeg)}, {\cellTwoRadius*cos(\resultAzimuthBeg)}) -- ({\cellTwoRadius*sin(\resultAzimuthEnd)}, {\cellTwoRadius*cos(\resultAzimuthEnd)}) -- ({\cellThreeRadius*sin(\resultAzimuthEnd)}, {\cellThreeRadius*cos(\resultAzimuthEnd)}) -- ({\cellThreeRadius*sin(\resultAzimuthBeg)}, {\cellThreeRadius*cos(\resultAzimuthBeg)});
    
    \path[fill=black] ({\cellFourRadius*sin(\resultAzimuthBeg)}, {\cellFourRadius*cos(\resultAzimuthBeg)}) -- ({\cellFourRadius*sin(\resultAzimuthEnd)}, {\cellFourRadius*cos(\resultAzimuthEnd)}) -- ({\cellFiveRadius*sin(\resultAzimuthEnd)}, {\cellFiveRadius*cos(\resultAzimuthEnd)}) -- ({\cellFiveRadius*sin(\resultAzimuthBeg)}, {\cellFiveRadius*cos(\resultAzimuthBeg)});
    
    \path[fill=red] ({\cellFiveRadius*sin(\resultAzimuthBeg)}, {\cellFiveRadius*cos(\resultAzimuthBeg)}) -- ({\cellFiveRadius*sin(\resultAzimuthEnd)}, {\cellFiveRadius*cos(\resultAzimuthEnd)}) -- ({\resultRadius*sin(\resultAzimuthEnd)}, {\resultRadius*cos(\resultAzimuthEnd)}) -- ({\resultRadius*sin(\resultAzimuthBeg)}, {\resultRadius*cos(\resultAzimuthBeg)});
    \draw[domain=\resultAzimuthBeg:\resultAzimuthEnd] plot ({\cellOneRadius*sin(\x)}, {\cellOneRadius*cos(\x)});
    \draw[domain=\resultAzimuthBeg:\resultAzimuthEnd] plot ({\cellTwoRadius*sin(\x)}, {\cellTwoRadius*cos(\x)});
    \draw[domain=\resultAzimuthBeg:\resultAzimuthEnd] plot ({\cellThreeRadius*sin(\x)}, {\cellThreeRadius*cos(\x)});
    \draw[domain=\resultAzimuthBeg:\resultAzimuthEnd] plot ({\cellFourRadius*sin(\x)}, {\cellFourRadius*cos(\x)});
    \draw[domain=\resultAzimuthBeg:\resultAzimuthEnd] plot ({\cellFiveRadius*sin(\x)}, {\cellFiveRadius*cos(\x)});
    \draw[domain=\resultAzimuthBeg:\resultAzimuthEnd] plot ({\resultRadius*sin(\x)}, {\resultRadius*cos(\x)});
    
    \end{tikzpicture}}}
    \subfloat[Add and delete.]{\resizebox{!}{0.15\textwidth}{\begin{tikzpicture}
    \draw (\rectangleOneBegX,\originalY) node[yshift=0.25cm,left] {$t_0$} rectangle +(\rectangleSize,\rectangleSize);
    \draw (\rectangleTwoBegX,\originalY) rectangle +(\rectangleSize,\rectangleSize);
    \draw[fill=black] (\rectangleThreeBegX,\originalY) rectangle +(\rectangleSize,\rectangleSize);
    \draw (\rectangleFourBegX,\originalY) rectangle +(\rectangleSize,\rectangleSize);
    \draw[fill=black] (\rectangleFiveBegX,\originalY) rectangle +(\rectangleSize,\rectangleSize);
    \draw (\rectangleSixBegX,\originalY) rectangle +(\rectangleSize,\rectangleSize);
    
    \draw (\rectangleSixEndX,\originalY) node[yshift=0.25cm,right] {$(1b)$};
    \draw[<->] (\rectangleOneBegX,{ \attackY + \rectangleSize / 2 }) node[left] {$t_1$} -- node [above, midway] {center bias} ({ \rectangleFiveBegX - 0.1 },{ \attackY + \rectangleSize / 2 });
    \draw[fill=red] (\rectangleFiveBegX,\attackY) rectangle +(\rectangleSize,\rectangleSize);
    \draw[fill=red] (\rectangleSixBegX,\attackY) rectangle +(\rectangleSize,\rectangleSize);

    \draw ({ \rectangleBegX + 6 * \rectangleSize },\attackY) node[yshift=0.25cm,right] {$(2b)$};
    \draw[<->] ({\rectangleBegX - 0.3}, { \originalY - 0.1 }) -- node [xshift=0.2cm,centered] {$\epsilon$} ({\rectangleBegX - 0.3}, { \attackY + \rectangleSize + 0.1 });
    \draw ({ \rectangleBegX + 6 * \rectangleSize },\resultY) node[yshift=0.6cm,right] {$(3b)$};
    \draw (0,0) -- ({ \resultRadius*sin(\resultAzimuthBeg) }, { \resultRadius*cos(\resultAzimuthBeg) });
    \draw (0,0) -- ({ \resultRadius*sin(\resultAzimuthEnd) }, { \resultRadius*cos(\resultAzimuthEnd) });
    \path[fill=red] ({\cellFourRadius*sin(\resultAzimuthBeg)}, {\cellFourRadius*cos(\resultAzimuthBeg)}) -- ({\cellFourRadius*sin(\resultAzimuthEnd)}, {\cellFourRadius*cos(\resultAzimuthEnd)}) -- ({\resultRadius*sin(\resultAzimuthEnd)}, {\resultRadius*cos(\resultAzimuthEnd)}) -- ({\resultRadius*sin(\resultAzimuthBeg)}, {\resultRadius*cos(\resultAzimuthBeg)});
    \draw[domain=\resultAzimuthBeg:\resultAzimuthEnd] plot ({\cellOneRadius*sin(\x)}, {\cellOneRadius*cos(\x)});
    \draw[domain=\resultAzimuthBeg:\resultAzimuthEnd] plot ({\cellTwoRadius*sin(\x)}, {\cellTwoRadius*cos(\x)});
    \draw[domain=\resultAzimuthBeg:\resultAzimuthEnd] plot ({\cellThreeRadius*sin(\x)}, {\cellThreeRadius*cos(\x)});
    \draw[domain=\resultAzimuthBeg:\resultAzimuthEnd] plot ({\cellFourRadius*sin(\x)}, {\cellFourRadius*cos(\x)});
    \draw[domain=\resultAzimuthBeg:\resultAzimuthEnd] plot ({\cellFiveRadius*sin(\x)}, {\cellFiveRadius*cos(\x)});
    \draw[domain=\resultAzimuthBeg:\resultAzimuthEnd] plot ({\resultRadius*sin(\x)}, {\resultRadius*cos(\x)});
    \end{tikzpicture}}}
    \caption{Hijacking techniques.}
    \label{fig:htech}
\end{figure}

%% file: sections/4-attack-description/section.tex
\section{Attack description}
\label{sec:radarh}
In this section, we detail the inner workings of the stealth malware that an adversary can exploit to execute an attack to a radar system.

\subsection{Overview}
\label{sec:radarh-overview}

\begin{figure*}
    \centering
    \includegraphics[width=\textwidth]{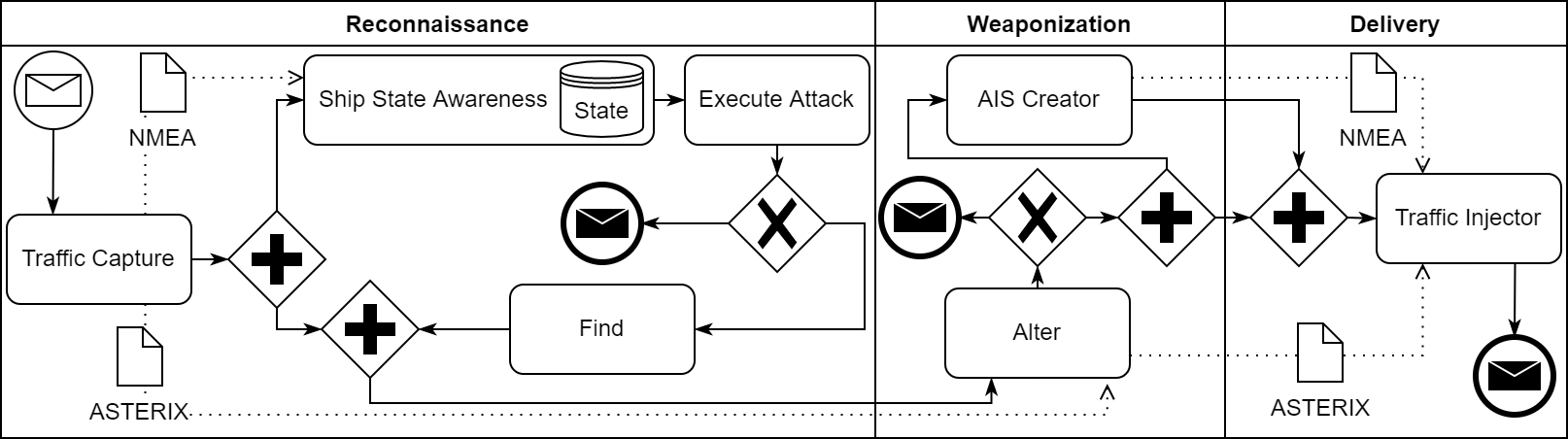}
    \caption{The workflow of the stealth malware.}
    \label{fig:attack-wf}
\end{figure*}

Figure~\ref{fig:attack-wf} shows the workflow of the stealth malware.
We map it to three steps that are inspired by the cyber kill chain~\cite{hutchins2011intelligence}.
In the \emph{reconnaissance} step, the malware evaluates its capabilities w.r.t. the current radar system (see Section~\ref{sec:radarh}), monitors the state of the ship, and determines whether an attack should start.
In the \emph{weaponization} step, the malware creates the ASTERIX and NMEA packets to alter the image on the PPI and updates the information on the INS devices, accordingly.
Finally, in the \emph{delivery} phase, the malware injects the weaponized packets into the bridge network.
We detail below the three steps.
\input{sections/4-attack-description/recon}

\input{sections/4-attack-description/weapon}
\input{sections/4-attack-description/exploit}

%% file: sections/4-attack-description/recon.tex
\subsection{Reconnaissance}
\label{sec:recon}

The \emph{reconnaissance} step starts with the \emph{traffic capture} task that captures the cleartext NMEA and ASTERIX traffic flowing into the bridge network.
Then, the \emph{ship state awareness} task analyzes the NMEA traffic to achieve the situation awareness.
The aim is twofold: detecting if the radar system under attack allows adversaries to apply the delete capability, and keeping updated the \emph{state} of the ship under attack.

Detecting the delete capability requires a single and short test of an attack after the malware starts.
At first, the malware listens for the \texttt{TTM} sentences from the ARPA system (see Section~\ref{sec:background-arpa}).
When it receives a \texttt{TTM}, it uses the position and bearing of the tracked target to execute a delete attack against the corresponding representation on the PPI.
After the attack starts, it waits for the time of nine consecutive scans of the PPI.
The ARPA system that stops sending \texttt{TTM}s for the target under attack or sends \texttt{TTM}s that contain a \emph{lost status}, means that the PPI grants the delete capability to the malware.

Updating the state requires reading data from NMEA sentences to track the ship telemetry (like position, speed, and bearing), nearby vessels, tracked targets, and weather and environmental conditions. 

The \emph{execute attack} task leverages the values of the above state to automate the decision to start the attack.
For example, the malware could trigger an attack at a specific GPS location, based on the position of nearby vessels, or if it is night.

If an attack requires operating in a specific area of the radar image, a \emph{find} task allows the malware to define the boundaries.
It uses a \emph{find} function that we detail below.

This step ends by forwarding the results of the find function and the captured ASTERIX packet to the \emph{weaponization} step.

\input{sections/4-attack-description/find}

%% file: sections/4-attack-description/find.tex
\subsubsection*{Find function}
\label{sec:find_function}
\input{sections/4-attack-description/find_annulus}
The find function returns a delimited zone of the radar image representing a given target, e.g., a ship or a waypoint on the ECDIS.

An example of such a zone is highlighted in Figure~\ref{fig:annulus}.
It is an annulus sector centered on point $O$ that has the latitude and longitude coordinates of the ship under attack, and contains the bounding box of the target (bbox).
Such a zone can be described with a tuple $\langle a_{min}, a_{max}, d_{min}, d_{max} \rangle$ where $a_{min},a_{max}$ and $d_{min},d_{max}$ are its ranges w.r.t. the angular and longitudinal dimensions, respectively.
\begin{algorithm}
\caption{The algorithm of the find function.}
\label{alg:find}
\begin{algorithmic}[1]
\Function{FIND}{$\rho$, $\theta$, $w$, $h$, $sm_\%$}
    \If{$\rho = 0$}
        \State \Return{$\langle0, 360, 0, \infty\rangle$}
    \EndIf
    \State $r \gets \frac{\sqrt{w^2+h^2}}{2}$
    \State $r^\star \gets r (1 + sm_\%)$
    \State $\phi = \atantwo({r^\star,\rho})$
    \State $a_{min/max} = C_{360}(\theta \pm \phi)$
    \State $d_{min} = \max \{ 0, \rho - r^\star \}$
    \State $d_{max} = d + r^\star$
    \State \Return{$\langle a_{min}, a_{max}, d_{min}, d_{max} \rangle$}
\EndFunction
\end{algorithmic}
\end{algorithm}

Algorithm~\ref{alg:find} represents the find function.
In the algorithm, $\rho$ is the distance between $O$ and the bbox center $(0, \infty)$, $\theta$ is the bearing of $O$ from the center of the bbox in arc degrees $[0, 360)$, $w$ and $h$ are the width and the height of the bbox $(0, \infty)$, and $sm_\%$ is the size margin of the bbox (Line 1). 

Since the find function is not always required, we consider $\rho = 0$ (Line 2) for returning a zone delimiting the entire radar image (Line 3).
Otherwise, we approximate the target shape with a circle inscribing its bbox.
A circle allows ignoring the orientation of the target rectangle during the calculation.
The circle has a radius $r$ corresponding to the half-diagonal of the bbox (Line 5).
The bounding circle can also be expanded by a percentage factor $sm_\%$, obtaining the final radius $r^\star$ (Line 6).
In this way, the buffer zone allows compensating position inaccuracies at the expense of a less precise find zone determination.
To calculate the angle span $\phi$ (Line 7), we observe that a right triangle exists between $O$, the center of the bounding circle, and one of the two points $P_{0/1}$ tangent to the circumference and passing through $O$.
Observing the symmetry of the problem concerning the vector joining $O$ and the center of the bounding circle, we can calculate the angular range $a_{min/max}$ (Line 8) where $C_{360}(x)$ represents the 360-degrees based complement of $x$.
Finally, knowing the distance $\rho$ from $O$ to the centroid, we can compute the range $d_{min/max}$ (Line 9) and return the tuple for the zone (Line 11).

\begin{example}
\label{ex:findship}
After having received an AIS message (see Section~\ref{sec:background-ais}) reporting the latitude ($lat_t$) and longitude ($lon_t$) of a target ship, we want to apply the find function for obtaining the tuple of its annulus section on the radar image.
To this aim, we must derive $O$, $\rho$, $\theta$, $w$, and $h$ parameters ($sm_\%$ is optional).

$O$ can be acquired from NMEA sentences generated by EPFS systems (see Section~\ref{sec:ship-navigation-network}), e.g., \texttt{GGA}, \texttt{GLL}, \texttt{GNS}, or \texttt{RMB} sentences.
We can obtain $\rho$ and $\theta$ by calculating the geodesic distance and azimuth, e.g., using Vincenty's inverse formula~\cite{vincenty}, between $O$ and  $\langle lat_{t}, lon_{t}\rangle$ and transforming
the resulting azimuth to the measuring ship's heading (\texttt{HDG}) relative azimuth.
\texttt{HDG} can be found from NMEA sentences originated by compasses or gyroscopes, e.g., \texttt{HDT} and \texttt{THS} sentences.
Finally, $w$ and $h$ can be obtained from AIS, i.e., \textit{ship static and voyage related data} messages that contain the size of the target ship.
\end{example}

%% file: sections/4-attack-description/find_annulus.tex
\begin{figure}[h]
\centering

\tikzmath{
    function bearingToReal(\x) {
        return 90 - \x;
    };
    \ownShipWidth = 0.5;
    \ownShipHeigth = 3 * \ownShipWidth;
    \otherShipWidth = \ownShipWidth;
    \otherShipHeigth = \ownShipHeigth;
    \otherShipBearing = 45;
    \otherShipDistance = 4;
    \otherShipOffsetX = \otherShipDistance * cos(bearingToReal(\otherShipBearing));
    \otherShipOffsetY = \otherShipDistance * sin(bearingToReal(\otherShipBearing));
    \otherShipRotation = 70;
    \otherShipCircleRadius = sqrt(\otherShipWidth * \otherShipWidth + \otherShipHeigth * \otherShipHeigth) / 2;
    \courseLineLength = 4;
    \bearingAngleLength = \courseLineLength * 0.5;
    \bearingAngleBegX = 0;
    \bearingAngleBegY = \bearingAngleLength;
    \bearingAngleMidX = \bearingAngleLength * cos(bearingToReal(\otherShipBearing / 2);
    \bearingAngleMidY = \bearingAngleLength * sin(bearingToReal(\otherShipBearing / 2);
    \bearingAngleEndX = \bearingAngleLength * cos(bearingToReal(\otherShipBearing);
    \bearingAngleEndY = \bearingAngleLength * sin(bearingToReal(\otherShipBearing);
    \angleSpan = atan(\otherShipCircleRadius / \otherShipDistance);
    \angleSpanMin = \otherShipBearing - \angleSpan;
    \angleSpanMax = \otherShipBearing + \angleSpan;
    \angleSpanMid = (\angleSpanMin + \otherShipBearing) / 2;
    \angleSpanLength = \courseLineLength * 0.65;
    \angleSpanBegX = \angleSpanLength * cos(bearingToReal(\angleSpanMin);
    \angleSpanBegY = \angleSpanLength * sin(bearingToReal(\angleSpanMin);
    \angleSpanMidX = \angleSpanLength * cos(bearingToReal(\angleSpanMid);
    \angleSpanMidY = \angleSpanLength * sin(bearingToReal(\angleSpanMid);
    \angleSpanEndX = \angleSpanLength * cos(bearingToReal(\otherShipBearing);
    \angleSpanEndY = \angleSpanLength * sin(bearingToReal(\otherShipBearing);
    \aMinLength = \courseLineLength * 0.25;
    \aMinMidAngle = \angleSpanMin * 0.5;
    \aMinBegX = \aMinLength * cos(bearingToReal(0);
    \aMinBegY = \aMinLength * sin(bearingToReal(0);
    \aMinMidX = \aMinLength * cos(bearingToReal(\aMinMidAngle);
    \aMinMidY = \aMinLength * sin(bearingToReal(\aMinMidAngle);
    \aMinEndX = \aMinLength * cos(bearingToReal(\angleSpanMin);
    \aMinEndY = \aMinLength * sin(bearingToReal(\angleSpanMin);
    \aMaxLength = \courseLineLength * 0.375;
    \aMaxMidAngle = \angleSpanMax * 0.5;
    \aMaxBegX = \aMaxLength * cos(bearingToReal(0);
    \aMaxBegY = \aMaxLength * sin(bearingToReal(0);
    \aMaxMidX = \aMaxLength * cos(bearingToReal(\aMaxMidAngle);
    \aMaxMidY = \aMaxLength * sin(bearingToReal(\aMaxMidAngle);
    \aMaxEndX = \aMaxLength * cos(bearingToReal(\angleSpanMax);
    \aMaxEndY = \aMaxLength * sin(bearingToReal(\angleSpanMax);
}
\begin{tikzpicture}
    \coordinate (ownShipCenter) at (0,0);
    \draw (ownShipCenter) node [yshift=-0.2 cm] {O};
    \coordinate (ownShipBLCorner) at ($ (ownShipCenter) - (\ownShipWidth / 2, \ownShipHeigth / 2) $);
    \coordinate (ownShipTRCorner) at ($ (ownShipCenter) + (\ownShipWidth / 2, \ownShipHeigth / 2) $);
    \draw (ownShipBLCorner) rectangle (ownShipTRCorner);
    \coordinate (courseLineEnd) at ($ (ownShipCenter) + (0, \courseLineLength) $);
    \draw[thin,dashed] (ownShipCenter) -- (courseLineEnd) node [midway, above, sloped] {\texttt{HDG}};
    \coordinate (otherShipCenter) at ($(ownShipCenter) + (\otherShipOffsetX, \otherShipOffsetY)$);
    \draw[thin,dashed,name path=bearingLine] (ownShipCenter) -- (otherShipCenter) node [pos=0.65, below, sloped] {$\rho$};
    \begin{scope}[rotate around={\otherShipRotation:(otherShipCenter)}]
        \coordinate (otherShipBLCorner) at ($ (otherShipCenter) - (\otherShipWidth / 2, \otherShipHeigth / 2) $);
        \coordinate (otherShipBRCorner) at ($ (otherShipCenter) + (\otherShipWidth / 2, 0) - (0, \otherShipHeigth / 2) $);
        \coordinate (otherShipTRCorner) at ($ (otherShipCenter) + (\otherShipWidth / 2, \otherShipHeigth / 2) $);
        \coordinate (otherShipPerpendicular90) at ($(otherShipCenter)!(otherShipBLCorner)!90:(ownShipCenter)$);
        \coordinate (otherShipPerpendicular270) at ($(otherShipCenter)!(otherShipTRCorner)!270:(ownShipCenter)$);
        \draw (otherShipBLCorner) rectangle (otherShipTRCorner);
    \end{scope}
    \draw (otherShipCenter) -- (otherShipBRCorner) node [below, pos=0.5] {r};
    \draw[thin,dashed,name path=otherShipBoundingCircle] (otherShipCenter) circle (\otherShipCircleRadius);
    \draw[thin,dashed] (otherShipPerpendicular90) -- (otherShipPerpendicular270);
    \draw (otherShipPerpendicular90) node [below] {$P_0$};
    \draw (otherShipPerpendicular270) node [xshift=-0.3 cm, yshift=0.15cm] {$P_1$};
    \coordinate (RAM1) at ($(otherShipCenter)!0.2!(otherShipPerpendicular90)$);
    \draw
        let
            \p1=(otherShipCenter),
            \p2=(RAM1),
        in
            coordinate (RAM2) at ($(otherShipCenter)!{veclen(\y2-\y1,\x2-\x1)}!(ownShipCenter)$);
    \coordinate (RAM3) at ($(RAM1)!(RAM2)!90:(otherShipCenter)$);
    \draw (otherShipCenter)
        -- (RAM1)
        -- (RAM3)
        -- (RAM2);
    \coordinate (otherShipBoundingCircleClosestPoint) at ($(otherShipCenter)!\otherShipCircleRadius cm!(ownShipCenter)$);
    \coordinate (otherShipBoundingCircleFarthestPoint) at ($(otherShipCenter)!-\otherShipCircleRadius cm!(ownShipCenter)$);
    \coordinate (minAngleLineBegPoint) at ($(otherShipPerpendicular270)!(otherShipBoundingCircleClosestPoint)!(ownShipCenter)$);
    \coordinate (maxAngleLineBegPoint) at ($(otherShipPerpendicular90)!(otherShipBoundingCircleClosestPoint)!(ownShipCenter)$);
    \coordinate (minAngleLineEndPoint) at ($(otherShipPerpendicular270)!(otherShipBoundingCircleFarthestPoint)!(ownShipCenter)$);
    \coordinate (maxAngleLineEndPoint) at ($(otherShipPerpendicular90)!(otherShipBoundingCircleFarthestPoint)!(ownShipCenter)$);
    
    \draw[thin,dashed, name path=minAngleLine] (ownShipCenter) -- (minAngleLineBegPoint);
    \draw[thin,dashed, name path=maxAngleLine] (ownShipCenter) -- (maxAngleLineBegPoint);
    
    \draw[line width=0.5mm] (minAngleLineBegPoint) -- (minAngleLineEndPoint);
    \draw[line width=0.5mm] (maxAngleLineBegPoint) -- (maxAngleLineEndPoint);
    \draw[line width=0.5mm] plot [smooth] coordinates {(minAngleLineBegPoint) (otherShipBoundingCircleClosestPoint) (maxAngleLineBegPoint)};
    \draw[line width=0.5mm] plot [smooth] coordinates {(minAngleLineEndPoint) (otherShipBoundingCircleFarthestPoint) (maxAngleLineEndPoint)};
    \coordinate (bearingIndicatorBeg) at ($(\bearingAngleBegX, \bearingAngleBegY)$);
    \coordinate (bearingIndicatorMid) at ($(\bearingAngleMidX, \bearingAngleMidY)$);
    \coordinate (bearingIndicatorEnd) at ($(\bearingAngleEndX, \bearingAngleEndY)$);
    \draw [->]
        (bearingIndicatorBeg)
        .. controls (bearingIndicatorMid) ..
        node [midway, above, sloped] {$\theta$}
        (bearingIndicatorEnd)
    ;
    \coordinate (angleSpanBeg) at ($(\angleSpanBegX, \angleSpanBegY)$);
    \coordinate (angleSpanMid) at ($(\angleSpanMidX, \angleSpanMidY)$);
    \coordinate (angleSpanEnd) at ($(\angleSpanEndX, \angleSpanEndY)$);
    \draw [<-]
        (angleSpanBeg)
        .. controls (angleSpanMid) ..
        node [midway, above, sloped] {$\phi$}
        (angleSpanEnd)
    ;
\end{tikzpicture}
\caption{An example annulus section and related quantities.}
\label{fig:annulus}

\end{figure}
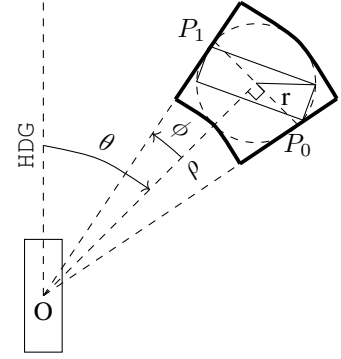

%% file: sections/4-attack-description/weapon.tex
\subsection{Weaponization}
\label{weapon}

This step starts by receiving return data from the find function and an ASTERIX packet.
The \emph{alter} task checks if the packet contains echoes related to the zone bounded by the find function.
If so, it calls an \emph{alter} function (see below) for creating a weaponized ASTERIX packet than can modify how such echoes appears on the PPI.

As the attack may involve adding ghost ships or altering the course and speed of existing targets, the malware can use the \emph{AIS creator} task that generates \texttt{VDM} sentences (see Section~\ref{sec:background-ais}) with data reflecting the changes occurring on the radar system.
Consequently, INS equipment using AIS will display information consistent with the attack.
This task aims at making the attack more difficult to detect since maritime operators can not rely on cross-checking procedures~\cite{Kristic2021}.
We detail below the alter function. 
\subsubsection*{Alter function}
\input{sections/4-attack-description/alter}

%% file: sections/4-attack-description/alter.tex
The alter function allows modifying echoes of an existing ASTERIX packet $Pkt$. 
Such an operation is performed by wrapping the execution of an user-provided variadic function $f : ( Pkt \times F_o \times F_a) \rightarrow Pkt \bigcup \emptyset$ where $Pkt$ is an existing ASTERIX packet, $F_o$ is the result of the \textit{find} function, and $F_a$ are user-specified arguments belonging to the domain of the function $f$.
Evaluation of alter returns the result of $f$, i.e., empty or an ASTERIX packet.

\begin{example}
\label{ex:copy-ship}
In Example~\ref{ex:findship}, we obtained the tuple of an annulus section related to a target ship.
Now, we want to copy its image into a different position, thus creating a \textit{ghost ship}.
We create a function \texttt{copy\_ship} that can be used with the alter function.
\begin{algorithm}
\small
\caption{The algorithm of \texttt{copy\_ship}.}
\label{alg:example-f}
\begin{algorithmic}[1]
\Function{copy\_ship}{$Pkt$, $a_{min}$, $a_{max}$, $d_{min}$, $d_{max}$, $o_a$, $o_d$}
    \If{$Pkt.start\_az < a_{min}$ \textbf{or} $Pkt.end\_az > a_{max}$}\label{alg:example-f:azimuth-check}
        \State \Return %
    \EndIf
    \State $i_o \gets $ \Call{Round}{$\frac{o_d}{Pkt.cell\_dur \cdot c/2}$}
    \State cells $\gets Pkt.cells$
    \State mod $\gets$ false %
    \For{$i \gets 0,Pkt.n\_cells$}
        \State $\rho_{min} \gets Pkt.cell\_dur \cdot (i + Pkt.center\_bias) \cdot \frac{c}{2}$\label{alg:example-f:cell-extent}
        \State $\rho_{max} \gets Pkt.cell\_dur \cdot (i + 1 + Pkt.center\_bias) \cdot \frac{c}{2}$
        \If{
            $\rho_{min} >= d_{min}$ \textbf{and} $\rho_{max} <= d_{max}$ \par\hskip\algorithmicindent
                \textbf{and} $i+i_o >= 0$ \textbf{and} $i+i_o < Pkt.n\_cells$ 
        } \label{alg:example-f:distance-check}
        \State $Pkt.cells[i+i_o] \gets cells[i]$ %
        \State mod $\gets$ true
        \EndIf
    \EndFor
    \If{mod} \label{alg:example-f:return-if-modified}
        \State $Pkt.start\_az \gets$ \Call{$C_{360}$}{$Pkt.start\_az + o_a$}
        \State $Pkt.end\_az \gets$ \Call{$C_{360}$}{$Pkt.end\_az + o_a$}
        \State \Return{$Pkt$} %
    \EndIf
    \State \Return %
\EndFunction
\end{algorithmic}
\end{algorithm}

Algorithm~\ref{alg:example-f} represents the implementation of the \texttt{copy\_ship} function.
In the algorithm, $Pkt$ is the original packet, $a_{min}$, $a_{max}$, $d_{min}$, $d_{max}$ are the values of the tuple, and $o_a$, $o_d$ are the angle and distance offsets at which the copy should be placed (Line 1).

The function begins by returning empty if the start and end angles included in the headers of $Pkt$, i.e, $Pkt.start\_az$ and $Pkt.end\_az$, are not in the angular range between $a_{min}$ and $a_{max}$ (Line 2-4).
Then, it calculates the cell index distance offset $i_o$ (Line 5) and copies the original video cell contents in a support variable (Line 6).
Following, for each cell in $Pkt$ (Line 8), it calculates the minimum $\rho_{min}$ (Line 9) and maximum $\rho_{max}$ (Line 10) covered distances as detailed in Section~\ref{sec:asterix}.
If a cell 1) has the covered distance included in the range between $d_{min}$ and $d_{max}$, and 2) copying its value would not exceed the bounds of the video block (Line 11), the algorithm copies the original cell value into the offset position (Line 12).
Finally, the function modifies the packet $Pkt$ azimuthal span returning it (Line 19) or empty if no modification happens (Line 21). 
\end{example}

%% file: sections/4-attack-description/exploit.tex
\subsection{Delivery}
\label{sec:exploit}

The \emph{delivery} step starts by receiving the weaponized NMEA and ASTERIX packets.
The \emph{traffic injector} task is in charge of injecting such packets into the navigation network.
As the involved protocols do not support authentication, this task forwards them to the multicast or broadcast addresses that INS equipment and the PPI use to communicate in the navigation network.
Once the above equipment consume the weaponized packets, they display the hijacked image and data. 

%% file: sections/5-attacks/section.tex
\section{Radar Hijacking}
\label{sec:attackimpl}

In this section, we will discuss two novel classes of attacks for radar hijacking leveraging the previous techniques.

\input{sections/5-attacks/dos}

\subsection{PPI poisoning attack}
\label{sec:misleading}

A PPI poisoning attack alters specific sections of the image shown on the PPI in real-time.
The aim is to induce the crew to make wrong decisions or to fail carrying out the required actions while underway.

This class of attacks is especially harmful during navigation in congested waters where the risk of collision is rather high.
The danger increases further in restricted visibility since navigation relies on the instruments under attack.

Ships in these risky situations avoid collisions by collectively interacting in accordance with COLREGs (see Section~\ref{sec:colregs}).
A radar under this attack may lead the victim to assess COLREGs with wrong assumptions.
In such a scenario, a vessel behaves differently than expected by others, and the risk of collisions remains high.

In the following, we show two implementations of this type of attack.
The first relies only on adding new echoes and can be executed on all radar systems, while the second requires a system granting the delete capability (see Section~\ref{sec:hijacking_tech}) to the attacker.

\input{sections/5-attacks/ghost}
\input{sections/5-attacks/hijack}

%% file: sections/5-attacks/dos.tex
\subsection{Denial of Service attack}
\label{sec:dos}

A Denial-Of-Service (DoS) attacks aims at rendering the radar system unusable and leaving the ship without means of safe navigation.
In this attack, the adversary overlays sectors or the entire azimuthal range of the radar image by filling them with echoes.
Below, we detail how adversaries can implement the different steps introduced in Section~\ref{sec:radarh}.

\subsubsection*{Reconnaissance}

We assume that a radar must continuously operate during the navigation.
For this reason, the malware does not need to implement specific checks during the \emph{execute attack} task.
Nevertheless, favorable conditions exist.
For example, they apply when vessels navigate in the darkness or congested areas.
They can be assessed by overhearing NMEA sentences with the current time and position and AIS information.

Since the attack corrupts the entire display, the \emph{find} task invokes the \emph{find} function with $\rho = 0$ as detailed in Section~\ref{sec:find_function}.

\subsubsection*{Weaponization}
In the field of network security, many DoS attacks rely on the misuse of protocols that accept small requests and amplify the volume of traffic to overwhelm a resource of the victim.
Protocols with a high amplification factor are the most effective for a DoS.
They require fewer resources to perform the attack and make adversaries harder to trace.

To execute a DoS against a radar system, the misuse of the ASTERIX protocol can enable a high amplification.
The angle span and the configurable number and duration of cells (see Section~\ref{sec:asterix}) are the amplification factors we use in the alter function of DoS attacks, namely the \emph{DoS} function.

\begin{algorithm}
\small
\caption{Denial of Service Attack.}
\label{alg:example-dos}
\begin{algorithmic}[1]
\Function{DoS}{$Pkt$, $a_{min}$, $a_{max}$, $d_{min}$, $d_{max}$, $i$, $k$}
 \State $Pkt.start\_az \gets 0$
 \State $Pkt.end\_az \gets 360$
 \State $n \gets \frac{32}{cell\_res}$
 \State $Pkt.n\_cells \gets n$
 \State $Pkt.cell\_dur \gets \frac{Pkt.cell\_dur \cdot Pkt.n\_cells}{n}$
 \State $Pkt.cells \gets [2^{cell\_res}, \ldots, 2^{cell\_res}]$
 \State $i \gets i + 1$
 \If{$i = k$}
    \State $\; i \gets 0$
    \State \Return $Pkt$
 \EndIf
 \State \Return
\EndFunction
\end{algorithmic}
\end{algorithm}

Algorithm~\ref{alg:example-dos} represents the implementation of the \emph{DoS} function. 
$i$ and $k$ are parameters used for controlling the injection rate as explained below.
The idea behind this function is to update the received packets with new ones containing echoes at maximum strength and covering the entire angle and distance span.

For covering the entire angle span, we set the \emph{start\_az} to 0 and \emph{end\_az} to 360 (Lines 2-3).
For the distance span, we use the minimum number $n$ of cells w.r.t. the constraints set by the ASTERIX protocol.
This solution creates a video block that is as small as possible.
To calculate $n$, we use the minimum number of bits in a video block, i.e., $32$, and the current cell resolution (Line 4-5).
As we replaced the number of cells in the original packet with $n$, we recalculate their \emph{cell\_dur} (Line 6). 
Then, we set each of the $n$ cells at the maximum strength (Line 7).

Since each altered packet covers a much greater angular span than the original one, adversaries can achieve the desired result without executing an injection at each received packet.
They can set $k$ to constrain that an attack happens once every $\frac{1}{k}$ legitimate packets.
A high value of $k$ further increases the amplification factor but widens the area of original image visible between each injection.

In the function, $i$ refers to a persistent counter which increments after each call (Line 8).
Once $i$ equals $k$, the modified packet is returned and $i$ is reset to 0 (Lines 10-11).
Otherwise, a null value is returned, indicating that no injection has to take place (Line 13).

This attack does not require creating new AIS sentences and the weaponization step ends without running the \emph{AIS creator task}. 
\subsubsection*{Delivery}
\begin{figure}[h]
    \centering
    \includegraphics[width=.30\textwidth]{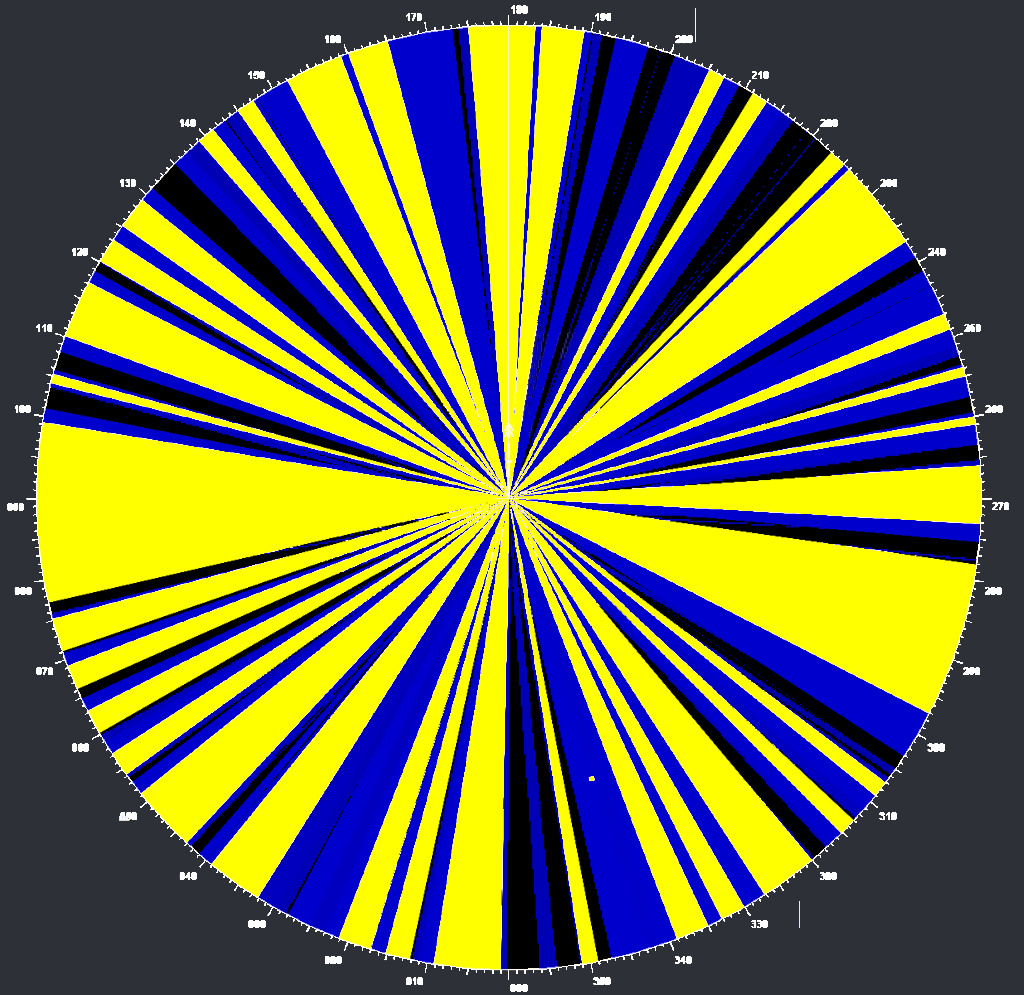}
    \caption{Radar after a DoS attack with $k=10$.}
    \label{fig:dos-after}
\end{figure}

Whenever the \emph{DoS} function returns a non-null value, the \emph{traffic injector} task transmits the modified packet to the multicast address allocated to the radar system.
In Figure~\ref{fig:dos-after}, we show how the PPI looks like after a DoS attack and using $k=10$.

%% file: sections/5-attacks/ghost.tex
\subsubsection{Ghost ship}
\label{sec:watering-hole:ghost}

A ghost ship is a fictitious target that this attack introduces into the radar image.
Such a target appears as changing in time by following a trajectory, i.e., a set of waypoints and speed pairs.
Below we describe each step of the attack.

\subsubsection*{Reconnaissance}

Malware can keep a list of trajectories in the \emph{state} database of the \emph{ship state awareness} task.
As an example, we consider the trajectory that is represented in Figure~\ref{fig:trajectory}a.
It comprises the three points $P_0$, $P_1$ and $P_2$ to be undertaken at a constant speed $S_0 = S_1 = S_2$.
Each point is specified in a polar coordinate system w.r.t. an origin point $O$.
\input{sections/5-attacks/trajectory_drawing}

We program our malware to use this trajectory to reproduce a COLREGs crossing condition (see Section~\ref{sec:colregs}) and force the victim to perform an unexpected evasive maneuver in a congested area.
To this aim, the \emph{execute attack} task overhears AIS and ARPA sentences and triggers the attack when $(i)$ at least $2$ ships within a $6 \; nm$ radius are present, and $(ii)$ no ships are already present in the starboard bow area of the victim.

Once triggered, the malware initializes the waypoints $P_0^\prime$, $P_1^\prime$, and $P_2^\prime$ to create the crossing situation as depicted in Figure~\ref{fig:trajectory}b.
It uses the victim position as the origin $O$ and obtains their absolute coordinates ($lat$,$lon$) via the application of a geodesic formula (e.g., the one in~\cite{vincenty}).

The \emph{find} task manages the evolution of the ghost ship's position along the initialized trajectory once every $\Delta t$ time.
In particular, it applies the system of equations detailed below to generate a realistic behavior.
\begin{align}
    \underline{x}(t + \Delta t) & = v(\underline{x}(t), \; COG(t), \; SOG(t) \cdot \Delta t) \\
    \Delta C(t) &= C(t) - COG(t) \\
    \raisetag{10pt}
    \omega(t) &= \Omega \cdot sgn \begin{cases}
        \Delta C(t) + 360 & \text{if } \Delta C(t) < -180 \\
        \Delta C(t) - 360 & \text{if } \Delta C(t) > 180 \\
        \Delta C(t)       & \textit{otherwise}
    \end{cases} \\
    COG(t+\Delta t) &= C_{360}(COG(t) + \omega(t) \cdot \Delta t) \\
    a(t) &= A \cdot sgn(S(t) - SOG(t)) \\
    SOG(t+\Delta t) &= SOG(t) + a(t) \cdot \Delta t
\end{align}
The task at a time $t_0$ initializes $\underline{x}$ to $P_0^\prime$, $C(t_0)$ and $COG(t_0)$ to the bearing between $P_0$ and $P_1$, $S(t_0)$ and $SOG(t_0)$ to $S_0$.
$C(t)$ changes according to the closest points of the trajectory.
$\Omega$ and $A$ are two constants constraining the maximum rotation speed and acceleration for the ghost ship.

The position $\underline{x}$ evolves according to the current course and speed by using a geodesic destination formula $v$ (Eq. 1).
Eq. 2 and Eq. 3 calculate the angular velocity $\omega$.
It is an on-off feedback control for the $COG$ variable w.r.t the target COG $C$. 
Then, $COG$ rotates according to $\omega$ (Eq. 4).
$C_{360}$ is the function detailed in Section~\ref{sec:find_function}.
The acceleration $a$ is an on-off feedback control for $SOG$ w.r.t. the target $S(t) = S_0$ (Eq. 5).
Finally, $SOG$ accelerates according to $a$ (Eq. 6).

The \emph{find} task ends by calling the find function with the $\underline{x}$ returned by the above system.

\subsubsection*{Weaponization}
In this step, the \emph{alter} task has to draw the ghost ship according to the annulus section returned by the \emph{find} function. It can leverage an implementation of the \emph{alter} function similar to Example~\ref{ex:copy-ship}.
For brevity, we omit how we generate the source image for the ghost ship.

Finally, the \emph{AIS creator} task uses $\underline{x}$, $COG$, and $SOG$ from the \emph{FIND} task to synthesize the corresponding \texttt{VDM} sentence.  

\subsubsection*{Delivery}
\begin{figure}[h]
    \centering
    \subfloat[\centering Real]{\includegraphics[width=0.242\textwidth]{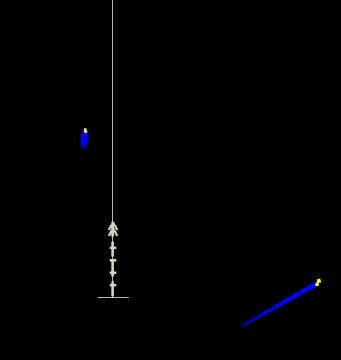}}
    \hfill
    \subfloat[\centering Attacked]{\includegraphics[width=0.242\textwidth]{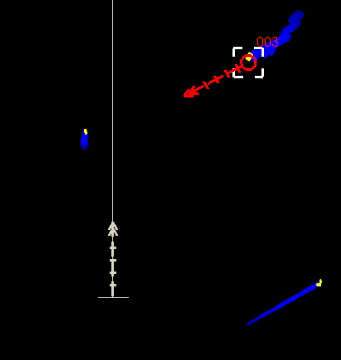}}
    \caption{Comparison of video feeds during~\ref{sec:watering-hole:ghost}.}
    \label{fig:ghost-ship}
\end{figure}

The \emph{traffic injector} injects weaponized AIS and ASTERIX packets.
NMEA devices show the position of the ghost ship as real ones.
PPIs display the video feed as in Figure~\ref{fig:ghost-ship}b instead of the real one as in Figure~\ref{fig:ghost-ship}a.
We set the PPI in head-up mode, and we enable trails (see Section~\ref{sec:radarsys}).
In both cases, the PPI displays two real targets, and in Figure~\ref{fig:ghost-ship}b, it also shows the ghost ship to the starboard bow of the victim.
In particular, the ghost ship's trail resembles the trajectory represented in Figure~\ref{fig:trajectory}b.
Moreover, the radar system acquires the ghost ship as a valid target, and ARPA marks it as a dangerous one (see Section~\ref{sec:background-arpa}).

The effects above show that the malware reproduced the conditions that lure operators to execute an evasive maneuver as desired.

%% file: sections/5-attacks/trajectory_drawing.tex
\tikzmath{
    \PZeroX = 1;
    \PZeroY = 2.8;
    \POneX = 1;
    \POneY = 1.8;
    \PTwoX = 0;
    \PTwoY = 1;
}
\begin{figure}[h]
    \centering
    \subfloat[Stored.]{%
        \resizebox{0.48\columnwidth}{!}{%
        \begin{tikzpicture}
        \tikzstyle{every node}=[font=\fontsize{12}{0}\selectfont]
        \begin{scope}[rotate around={270:(0,0)}]
            \draw[dotted] (0,0) -- (\PZeroX,\PZeroY);
            \draw[-] (0,0) -- (\POneX,\POneY) node [midway, below, sloped, rotate=270] {$\rho_1$};
            \draw[dotted] (0,0) -- (\PTwoX,\PTwoY);
            
            \draw[->] (0, 1.5) .. controls ({ 1.5 * sin(15) }, { 1.5 * cos(15) }) .. node [midway, right, yshift=+0.1cm] {$\theta_1$} ({ 1.5 * sin(28) }, { 1.5 * cos(28) });
            
            \draw (0,0) node [circle,draw,fill=black,inner sep=0pt, outer sep=0pt,minimum size=0.1cm] {} node [above] {$O$};
            \draw (\PZeroX,\PZeroY) node [circle,draw,fill=black,inner sep=0pt, outer sep=0pt,minimum size=0.1cm] {} node [below] {$P_0$};
            \draw (\POneX,\POneY) node [circle,draw,fill=black,inner sep=0pt, outer sep=0pt,minimum size=0.1cm] {} node [below] {$P_1$};
            \draw (\PTwoX,\PTwoY) node [circle,draw,fill=black,inner sep=0pt, outer sep=0pt,minimum size=0.1cm] {} node [above] {$P_2$};
        \end{scope}
        \end{tikzpicture}
        }
    }
    \subfloat[Initialized.]{%
        \raisebox{5mm}{
        \resizebox{0.50\columnwidth}{!}{%
        \begin{tikzpicture}
        \begin{scope}[rotate around={270:(0,0)}]
            \draw ({ \PTwoX - 0.25 },-1.0) rectangle ({ \PTwoX + 0.25 },-0.25) node[pos=.5] {$O$};
            \draw ({ \PZeroX + 0.25 },\PZeroY) rectangle ({ \PZeroX - 0.25 },{ \PZeroY + 0.75 });
            \draw[dotted] (0,-0.25) node [above, xshift=0.55cm] {Victim} -- (0, 1);
            \draw[->, thick]
                   (\PZeroX, \PZeroY) node [below, xshift=-0.5cm] {Ghost} node [above, xshift=-0.23cm] {$P_0^\prime$}
                -- (\POneX, \POneY) node [above, xshift=0.1cm] {$P_1^\prime$}
                -- (\PTwoX, { \PTwoY + 0.1 }) node [right, xshift=-0.1cm, yshift=0.2cm] {$P_2^\prime$};
        \end{scope}
        \end{tikzpicture}
        }
        }
    }
    \caption{An example trajectory comprised of three points.}
    \label{fig:trajectory}
\end{figure}

%% file: sections/5-attacks/hijack.tex
\subsubsection{Ship trajectory hijack}
\label{sec:watering-hole:trajectory-hijack}
A ship trajectory hijack exploits the delete capability to modify the trajectory of an existing target in the radar image.

As an example, we consider the victim in an overtaking situation (see Section~\ref{sec:colregs}).
The adversaries aim at modifying the trajectory of the vessel being overtaken so that no evasive maneuvers seem to be required.

The steps of the attack can be outlined as deleting the real target's echo and adding a ghost ship with the new trajectory, as described below.
\subsubsection*{Reconnaissance}
The \emph{execute attack} task overhears AIS and ARPA sentences and triggers the attack when a target $(i)$ goes at a slower speed w.r.t. the victim, and $(ii)$ has an angle between its beam and the victim bow of at least $22.5^\circ$.
After triggering, the attack initializes a hijacked trajectory $T$ as the one depicted in Figure~\ref{fig:trajectory}, but exchanges the order between $P_0$ and $P_2$.
$T$ has speeds $S_0 = S_1 = S_2$ set to a value exceeding the victim's one.

The \emph{find} task executes two find operations.
The first returns the annulus section of the overtaken ship as the one described in Example~\ref{ex:findship}.
The second follows the implementation as in the ghost ship attack and using $T$.

\subsubsection*{Weaponization}
This step invokes two implementations of the alter function according to the results of the two find functions above.
The first takes the annulus section of the overtaken ship as input and deletes its echoes.
Its implementation relies on setting the echo strengths to 0 in the annulus section and altering the \emph{center\_bias} value to force the PPI to replace echoes (see Section~\ref{sec:hijacking_tech}).
The second follows the implementation as in the ghost ship attack.

Finally, the \emph{AIS Creator} task generates \texttt{VDM} sentences according to the modified trajectory.

\subsubsection*{Delivery}
During this attack, the weaponized AIS messages have to coexist with the real ones.
A solution to make the malicious one prevail needs that the \emph{traffic injector} task injects them at a time interval less than $2$s, i.e., less than the one set in the standard (see~\cite{ais}).

\begin{figure}[h]
    \centering
    \subfloat[\centering Real]{\includegraphics[width=0.243\textwidth]{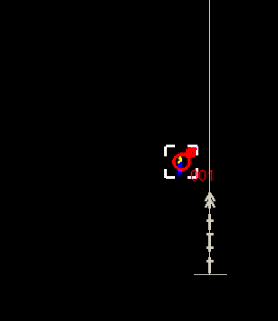}}
    \hfill
    \subfloat[\centering Attacked]{\includegraphics[width=0.243\textwidth]{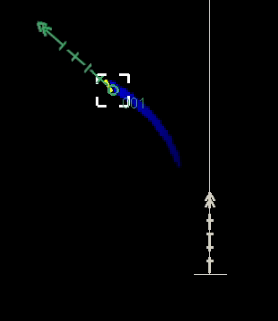}}
    \caption{Comparison of video feeds during~\ref{sec:watering-hole:trajectory-hijack}.}
    \label{fig:attack-hijack}
\end{figure}

Poisoned PPIs display the video feed as in Figure~\ref{fig:attack-hijack}b instead of the real one as in Figure~\ref{fig:attack-hijack}a.
Although the ARPA marks the overtaken ship as dangerous, the victim appears out of a safe distance in the real scenario.
In the attacked scenario, PPI shows the overtaken ship performed a maneuver that led it to get out of the overtaking situation.

Again, the malware creates the conditions to lure the victim as desired.

%% file: sections/6-detection/section.tex
\section{Detection}
\label{sec:detection}
In this section, we detail the design and the implementation of a detection system for the previously described attacks.

\subsection{Overview}
\label{sec:detectoverview}
As previously mentioned, performances of radar equipment must comply with standards and regulations established by IMO.
Moreover, the operation strictly follows the manufacturer specifications, e.g., resolution or speed of antennas, and depends on onboard configurations, e.g., SIC/SAC or IP addresses, that do not change over time.
As a result, a list of rules that constrain standards and regulations, manufacturer specifications, and onboard configurations can determine the expected behavior of a radar system.  

For this reason, we design the detection solution as a policy enforcement system where policies define the conditions under which a radar system is operating as expected.
The above policies can be expressed on values, their calculated aggregations, e.g.,  mean or variance, or frequency distribution obtained from the information carried by ASTERIX packets.
To keep the solution as much general as possible, it takes as input \emph{candidate policies} (see Section~\ref{sec:policy_generator}).
A candidate policy contains conditions that specify its eligibility for the radar system under monitoring and uses variables to refer to quantities that depend on single manufacturers or onboard configurations.
Our solution automatically infers the eligibility of candidate policies and the values of their variables after it receives a proper amount of ASTERIX traffic.

This implementation provides two main benefits: first, it can automatically tailor to every ship configuration; second, it can detect all the attacks that aim at violating the normal operation of a radar system since it models the expected behavior in any running configuration. 
Moreover, our solution operates by connecting to the bridge network and listening for the multicast traffic like the other INS equipment.
For this reason, it does not require onboard systems redesign, standardization and certification.

\begin{figure*}
    \centering
    \includegraphics[width=\textwidth]{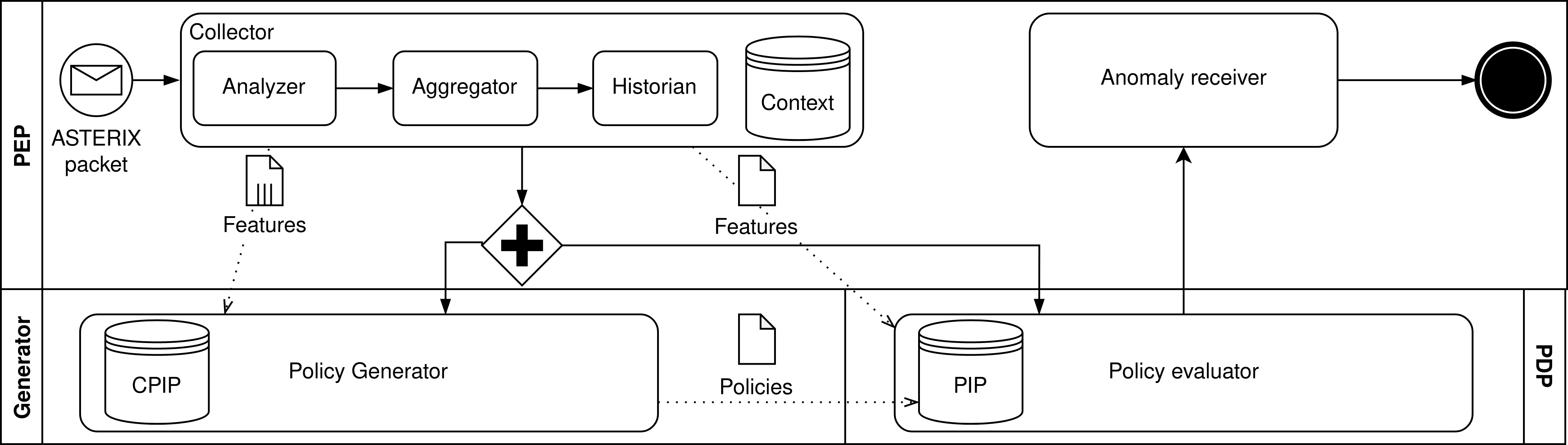}
    \caption{The workflow of the detection system.}
    \label{fig:detection}
\end{figure*}

In Figure~\ref{fig:detection}, we depict the workflow of our detection solution.
Next, we present each task in detail.

\subsection{Collector}
\label{sec:collector}

The \emph{collector} implements packets capture and analysis as part of our system Policy Enforcement Point (PEP) functionality.
It connects to the bridge network, receives the ASTERIX traffic via multicast or broadcast, and has preconfigured the unique IDs of antennas admitted to transmitting data to the PPI.
The collector relies on three components: the \emph{analyzer}, the \emph{aggregator}, and the \emph{historian}.

The \emph{analyzer} parses each packet and returns the unique ID of the sender antenna, and values from CAT-240 header (see Section~\ref{sec:asterix}), e.g., center bias or covered distance.

The \emph{aggregator} keeps a buffer of past values that the analyzer returns and calculates their aggregations, e.g., mean and variance, or frequency distribution.
Aggregations occur after each revolution.

The \emph{historian} returns a time series of values of the aggregator using data stored in the context database.

As a result, the collector creates the \emph{feature} $F$ of the received packet.
$F$ is a tuple $\langle S, D, A, A^n \rangle$ where $S$ has the unique ID of the sender antenna, $D$ has the data from the analyzer, $A$ has the quantities produced by the aggregator, and $A^n$ is the time series produced by the historian.
The current $F$ is stored in the context database that keeps the last available features for every $S$.

The task ends by forwarding $F$ to the \emph{policy evaluator}, and a set $F_g$ consisting of $F$ and the latest stored feature for each subject $s \neq F.S$ present in the context database to the \emph{policy generator}.
\subsection{Policy generator}
\label{sec:policy_generator}

The \emph{policy generator} relies on a list of candidate policies $P_c$ stored in the CPIP database and the set of features $F_g$ received from the \emph{collector}.
It can generate the policies to be applied by the \emph{policy evaluator} in response to the observed input data.
\emph{Candidate} policies $P_c$ are tuples $\langle P_a, T\rangle$.
Within $P_c$, the \emph{activation policy} $P_a$ is a function $F_g \rightarrow \mathbb{B} \bigcup \{undecided\}$ used to determine if a given $P_c$ is applicable to the current system configuration.
$T: F_g \rightarrow P_f$ is a \emph{transformation function} that takes as input a set of features $F_g$ and returns a \emph{policy evaluator} compatible policy $P_f : F \rightarrow \mathbb{B} \bigcup \{undecided\}$.
When the policy generator receives a feature set $F_g$ from the collector, it evaluates the $P_a$ associated with each candidate policy.
Evaluation of $P_a$ to \emph{false} signals that $P_c$ has been deemed incompatible with the observed data.
Conversely, for a \emph{true} verdict, $T$ is evaluated with the same argument as $P_a$ to generate a policy $P_f$ that is subsequently transferred to the PIP database.
While undecided results are ignored, boolean verdicts also result in the removal of the examined $P_c$ from the CPIP.

To clarify the process of policies generation, we propose the following example.
\input{sections/6-detection/sic-sac}

\subsection{Policy evaluator and anomaly receiver}
The \emph{policy evaluator} implements our system's Policy Decision Point (PDP) functionality.
It evaluates policies in the PIP against the features $F$ received from the collector.
If any policy violation occurs, it returns an anomaly containing the description of the violated policy and the feature that triggered it.

Finally, the \emph{anomaly receiver} implements the functionality of PEP that enforces PDP decisions.
In particular, it collects anomalies and executes an action accordingly.
For instance, it might operate by generating alerts targeted at the bridge alert management systems~\cite{bridgeAlert}, or by feeding dedicated solutions as we proposed in our implementation.
\input{sections/6-detection/implementation}

%% file: sections/6-detection/sic-sac.tex
\begin{example}
\label{ex:azspan-policy}
According to international standards~\cite{radarPerformance81,radarTesting}, an antenna should scan clockwise, continuous, and automatic through  $\ang{360}$ of azimuth.
To this aim, in a single antenna configuration, we want to create a candidate policy for imposing the azimuthal span to lie within three standard deviations w.r.t. its estimated mean, i.e., the 68-95-99.7 rule.
Each element of $A^n$ contains, among others, the azimuthal span sample mean $\mu_{az}$, and the biased sample variance of the azimuthal span $s_{az}$.
Generating an applicable policy depends on a reasonable estimation of the mean and standard deviation parameters.
A possible heuristic is imposing the sample variance of the observed means and variances to be below some thresholds $\alpha$ and $\beta$.
An activation policy $P_a$ matching this description, characterized by the design parameters $\alpha$ and $\beta$, is
$$
P_a = Var(\underline{\mu_{az}}) \leq \alpha \bigwedge Var(\underline{s_{az}}) \leq \beta
$$
Upon the triggering of $P_a$, evaluation of the transformation function $T$ will produce a policy $P_f$.
$$
    T(F) = P_f = (\overline{\mu_{az}} - 3 \cdot \sigma_{az}) \leq \mu_{az} \leq (\overline{\mu_{az}} + 3 \cdot \sigma_{az})
$$
Such policy will consist of a single clause enforcing the azimuthal span value to be between $\mu - 3\sigma$ and $\mu + 3\sigma$, where $\mu$ and $\sigma$ are the mean and standard deviation obtained from the samples which triggered $P_a$.
\end{example}

%% file: sections/6-detection/implementation.tex
\subsection{Implementation}
\label{sec:detection_implementation}

We realized our detection system using the Rust programming language~\cite{Matsakis2014} for implementing core tasks, Open Policy Agent (OPA)~\cite{OpenPolicyAgent} for the policy engine, and Lua~\cite{Ierusalimschy1996} as the scripting language for defining transformation functions of \emph{candidate} policies.

The system's configuration requires the list of unique IDs of transmitting antennas (i.e., the union of their SIC/SAC, IP address, and the network port used for communicating with PPI), the IP multicast address to bind, and where to forward anomalies.

The main thread starts capturing traffic received through the multicast.
Then, it performs analysis on each packet to extract and calculate its features.

Values from features represent inputs for threads implementing the functionalities of the \emph{policy generator} and \emph{evaluator}.
As detailed in Section~\ref{sec:policy_generator}, the \emph{policy generator} receives a list of features related to the last packet and the last ones collected from the other antennas in the system.

An OPA instance evaluates \emph{activation} policies from available \emph{candidate} policies from a repository directory.
Policies are expressed in Rego, i.e., the declarative language of OPA.

After an \emph{activation} policy is admissible, the \emph{policy generator} executes the embedded interpreter to evaluate the Lua script generating the entry for the PIP database.

The thread implementing the functionalities of the \emph{policy evaluator} interacts with the OPA instance, evaluating each entry stored in the PIP against the features received from the collector. 
If the OPA instance returns an anomaly, it forwards its details to a web application that acts as our \emph{anomaly receiver}.

\begin{figure}[h]
    \centering
    \includegraphics[width=.5\textwidth]{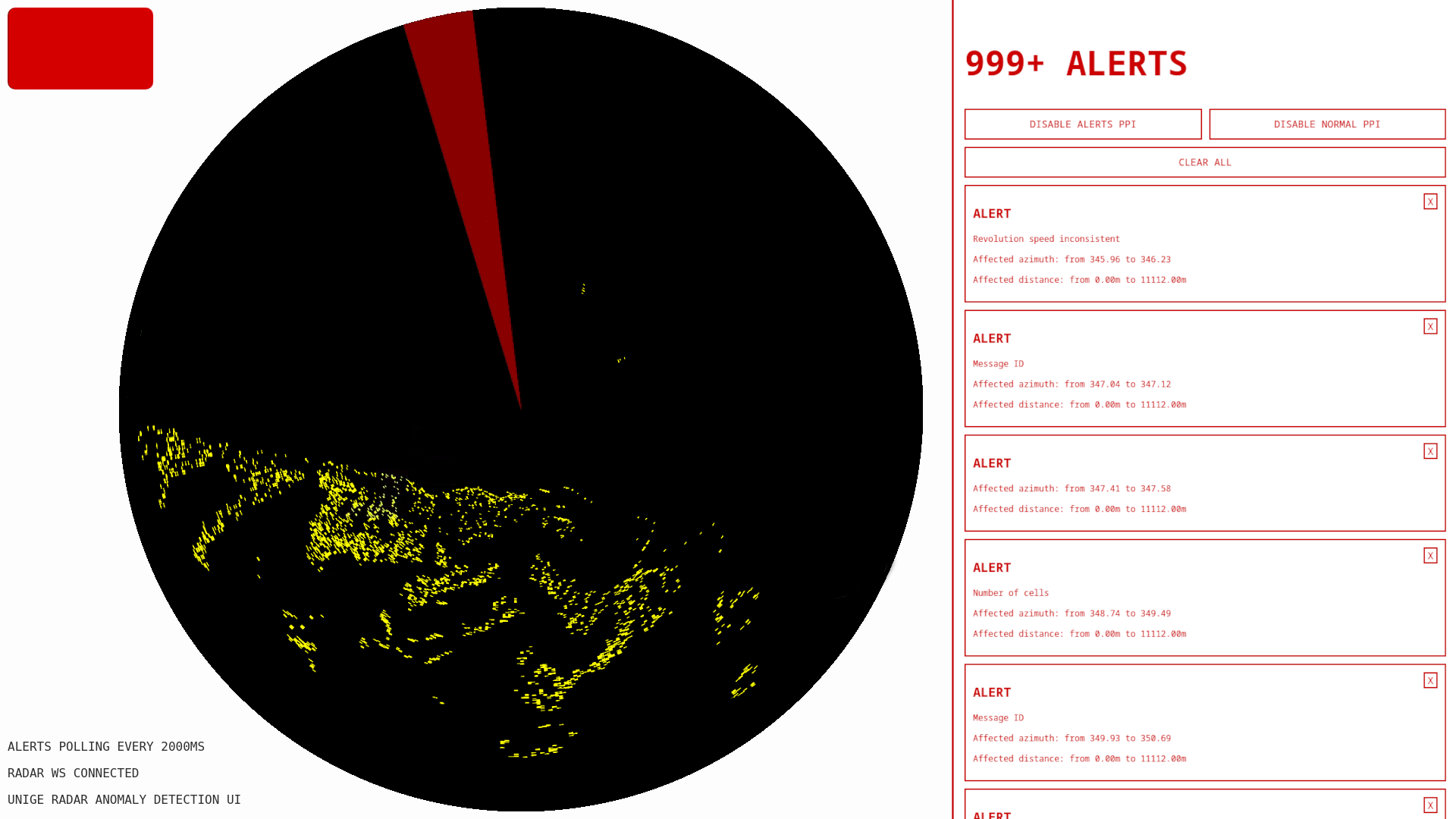}
    \caption{Anomalies shown during~\ref{sec:watering-hole:trajectory-hijack}.}
    \label{fig:detection-ppi}
\end{figure}

The \emph{anomaly receiver} appears as a secondary PPI.
When it receives an anomaly, it shows an acknowledgeable alarm and highlights what sectors of the radar image are affected by the anomaly.
Figure~\ref{fig:detection-ppi} shows the anomalies produced by the attack in~\ref{sec:watering-hole:trajectory-hijack}.

%% file: sections/7-evaluation/section.tex
\section{Experimental evaluation}
\label{sec:expeval}
In this section, we demonstrate the practical feasibility of the attacks against a radar system, and evaluate our detection system during their execution.

\input{sections/7-evaluation/performance_tables}
\begin{table*}[!t]
    \begin{minipage}{.5\linewidth}
        \input{sections/7-evaluation/detection_results_table}
    \end{minipage}
    \begin{minipage}{.5\linewidth}
        \input{sections/7-evaluation/policies_table}
    \end{minipage}
\end{table*}

\begin{figure*}[t]
    \centering
        \input{sections/7-evaluation/tcpa_ghost}
    \hfill
        \input{sections/7-evaluation/tcpa_overtaking}
    \caption{Evolution of the TCPA during attacks~\ref{sec:watering-hole:ghost} (left) and \ref{sec:watering-hole:trajectory-hijack} (right).}
    \label{fig:attack-tcpa-behavior}
\end{figure*}

\input{sections/7-evaluation/setting}

\input{sections/7-evaluation/results}

\input{sections/7-evaluation/discussion}

%% file: sections/7-evaluation/performance_tables.tex
\begin{table*}[!t]
    \begin{minipage}{.6\linewidth}
\caption{Attack performances.}
\centering
\begin{tabular}
{c |                                       r r r r r|                                                 r r|                                          r r| }
 \multirow{3}{*}{Attack}                                    & \multicolumn{5}{c|}{ASTERIX Packets {\scriptsize (MiB)}}         & \multicolumn{2}{c|}{CPU {\scriptsize (\%)}} & \multicolumn{2}{c|}{RAM {\scriptsize (B)}} \\
 \cline{2-10}
                                           & \multicolumn{2}{c}{Legitimate} & \multicolumn{2}{|c|}{Attacker} & \multicolumn{1}{c|}{\multirow{2}{*}{$\frac{\sum A}{\sum (L+ A)}$}} & & & & \\
                                           & \multicolumn{1}{c}{$\mu$} & \multicolumn{1}{c}{$\sigma$} & \multicolumn{1}{|c}{$\mu$} & \multicolumn{1}{c|}{$\sigma$} & & \multicolumn{1}{c}{$\mu$} & \multicolumn{1}{c|}{$\sigma$} & \multicolumn{1}{c}{$\mu$} & \multicolumn{1}{c|}{$\sigma$} \\
 \hline \hline
 \ref{sec:dos}                             & 32.90  & 8.01     & 0.04  & 0.0004   & 0.012\%            & 3.660 & 1.595                                             & 3261  & 147.8 \\
 \ref{sec:watering-hole:ghost}             & 35.37  & 1.66     & 0.05  & 0.009    & 0.144\%            & 3.934 & 1.497                                             & 3270  &  96.8 \\
 \ref{sec:watering-hole:trajectory-hijack} & 104.99 & 3.26     & 1.22  & 0.25     & 1.148\%            & 4.226 & 1.575                                             & 3323  & 131.3 \\
\end{tabular}
\label{tab:attack-performance}

     \end{minipage} 
     \begin{minipage}{.4\linewidth}

\caption{Detection performances.}
\centering
\begin{tabular}{c| r r| r r|}
\multirow{2}{*}{Attack} & \multicolumn{2}{c}{CPU (\%)} & \multicolumn{2}{c}{RAM (KiB)} \\
\cline{2-5}
 & \multicolumn{1}{c}{$\mu$} & \multicolumn{1}{c|}{$\sigma$} & \multicolumn{1}{c}{$\mu$} & \multicolumn{1}{c|}{$\sigma$} \\
\hline\hline
\ref{sec:dos}                             &  8.830 & 2.605 & 181.00 & 1.350 \\
\ref{sec:watering-hole:ghost}             & 10.304 & 4.645 & 181.51 & 1.187 \\
\ref{sec:watering-hole:trajectory-hijack} & 17.681 & 2.937 & 584.87 & 3.540 \\
\end{tabular}
\label{tab:detection-performance}
     \end{minipage} 
\end{table*}

%% file: sections/7-evaluation/detection_results_table.tex
\caption{Detection results.}
\label{tab:detection_res}
\centering
\begin{tabular}
 {c |r r| r r r r}
 Attack & \multicolumn{2}{c|}{Packets} & \multicolumn{4}{c|}{Detection} \\
 \hline
 & \multicolumn{1}{c}{Legit} & \multicolumn{1}{c|}{Attack} & \multicolumn{2}{c}{True positive} & \multicolumn{2}{c|}{False positive} \\
 \hline \hline
 ~\ref{sec:dos}                             & 816676  & 29012 & 29012 & (100.0\%) & 967  & (0.114\%) \\
 ~\ref{sec:watering-hole:ghost}             & 877957  & 1266  & 1256  & (99.21\%) & 900  & (0.102\%) \\
 ~\ref{sec:watering-hole:trajectory-hijack} & 2606273 & 30266 & 30266 & (100.0\%) & 2340 & (0.088\%) \\
\end{tabular}

%% file: sections/7-evaluation/policies_table.tex
\caption{Detection rate of malicious packets for each policy.}
\label{tab:matched-policies}
\centering
\begin{tabular}
 {c | r | r | r | r | r | r |}
 Attack                                     & \multicolumn{1}{c|}{$P_1$ {\scriptsize (\%)}} & \multicolumn{1}{c|}{$P_2$ {\scriptsize (\%)}} & \multicolumn{1}{c|}{$P_3$ {\scriptsize (\%)}} & \multicolumn{1}{c|}{$P_4$ {\scriptsize (\%)}} & \multicolumn{1}{c|}{$P_5$ {\scriptsize (\%)}} & \multicolumn{1}{c|}{$P_6$ {\scriptsize (\%)}} \\
 \hline\hline
 ~\ref{sec:dos}                             & {\color{darkgray}0.000} & {\color{darkgray}0.000} & 100.0 & 0.103 & {\color{darkgray}0.000}  & {\color{darkgray}0.000} \\
 ~\ref{sec:watering-hole:ghost}             & {\color{darkgray}0.000} & {\color{darkgray}0.000} & {\color{darkgray}0.000} & {\color{darkgray}0.000} & 99.21 & {\color{darkgray}0.000} \\
 ~\ref{sec:watering-hole:trajectory-hijack} & 100.0 & 100.0 & 20.43 & 0.135 & 100.0 & {\color{darkgray}0.000} \\
\end{tabular}

%% file: sections/7-evaluation/tcpa_ghost.tex
\begin{tikzpicture}
\begin{axis}[
    height=4cm, width=.48\textwidth,
    xlabel={Time [s]},
    ylabel={TCPA [m]},
    xmin=0, xmax=120,
    ymin=15, ymax=19,
    legend pos=north west,
    ymajorgrids=true,
]

\addplot+ [
  blue, smooth, mark=o
] coordinates {
 (0.0, 17.954054054054055) (20.0, 17.804081632653062) (40.0, 17.392) (60.0, 16.86082474226804) (80.0, 16.4578125) (100.0, 15.89) (120.0, 15.1)
};
\addplot+ [
  smooth, name path=A, mark=none, draw=none
] coordinates {
 (0.0, 18.583293470821214) (20.0, 18.263107577464112) (40.0, 17.953505911713215) (60.0, 17.34399066895863) (80.0, 16.888686358276445) (100.0, 16.41799410771123) (120.0, 15.1)};
\addplot+ [
  smooth, name path=B, mark=none, draw=none
] coordinates {
 (0.0, 17.324814637286895) (20.0, 17.345055687842013) (40.0, 16.830494088286784) (60.0, 16.37765881557745) (80.0, 16.026938641723554) (100.0, 15.36200589228877) (120.0, 15.1)};

\addplot [blue!25] fill between [ of=B and A ];

\end{axis}
\end{tikzpicture}

%% file: sections/7-evaluation/tcpa_overtaking.tex
\begin{tikzpicture}
\begin{axis}[
    height=4cm, width=.48\textwidth,
    xlabel={Time [s]},
    ylabel={TCPA [m]},
    xmin=0, xmax=600,
    ymin=-20, ymax=45,
    legend pos=north west,
    ymajorgrids=true,
]

\addplot+ [
  blue, smooth, mark=o
] coordinates {
 (0.0, 33.546) (20.0, 33.38) (40.0, 33.025999999999996) (60.0, 26.516) (80.0, 21.899) (100.0, 22.675) (120.0, 27.387) (140.0, 39.973) (160.0, 41.234) (180.0, 41.298) (200.0, 42.169) (220.0, 40.832) (240.0, 30.549) (260.0, 17.861) (280.0, 8.723) (300.0, 0.18799999999999992) (320.0, -13.085) (340.0, -15.62) (360.0, -14.872) (380.0, -14.903) (400.0, -14.641) (420.0, -13.836) (440.0, -13.469) (460.0, -12.189) (480.0, -10.304) (500.0, -8.708) (520.0, -8.461) (540.0, -8.383) (560.0, -8.363) (580.0, -8.272916666666667) (600.0, -8.251041666666667) (620.0, -8.107777777777777) (640.0, -7.550666666666666) (660.0, -7.35) (680.0, -5.4)
};
\addplot+ [
  smooth, name path=A, mark=none, draw=none
] coordinates {
 (0.0, 44.156455112) (20.0, 43.84508997133365) (40.0, 42.07641569445427) (60.0, 33.486786370723365) (80.0, 26.981004197820646) (100.0, 27.27544738711532) (120.0, 38.168209382883525) (140.0, 68.90924428520974) (160.0, 72.89692233046486) (180.0, 71.36272136420145) (200.0, 64.47701683591607) (220.0, 67.13653692096219) (240.0, 55.92506242986576) (260.0, 41.603158280113874) (280.0, 32.273795542976266) (300.0, 16.361005817463777) (320.0, -2.8959944875207118) (340.0, -8.790133917339908) (360.0, -10.8997202942635) (380.0, -11.241834567622993) (400.0, -11.777820478961026) (420.0, -10.883530737251487) (440.0, -10.690304171908968) (460.0, -8.98156246271957) (480.0, -7.580999575260535) (500.0, -6.829806154516493) (520.0, -6.989252944987015) (540.0, -7.1094893824827) (560.0, -7.157151038822972) (580.0, -7.1019989728836945) (600.0, -7.124987092718863) (620.0, -6.8657037802917324) (640.0, -5.8816020853587485) (660.0, -5.689455689417651) (680.0, -5.4)};
\addplot+ [
  smooth, name path=B, mark=none, draw=none
] coordinates {
 (0.0, 23.122155698) (20.0, 22.914910028666355) (40.0, 23.975584305545723) (60.0, 19.54521362927663) (80.0, 16.816995802179356) (100.0, 18.07455261288468) (120.0, 16.605790617116476) (140.0, 11.03675571479026) (160.0, 9.571077669535136) (180.0, 11.233278635798559) (200.0, 19.86098316408392) (220.0, 14.527463079037819) (240.0, 5.172937570134241) (260.0, -5.881158280113873) (280.0, -14.827795542976263) (300.0, -15.985005817463778) (320.0, -23.27400551247929) (340.0, -22.44986608266009) (360.0, -18.8442797057365) (380.0, -18.56416543237701) (400.0, -17.504179521038974) (420.0, -16.788469262748514) (440.0, -16.247695828091032) (460.0, -15.39643753728043) (480.0, -13.027000424739466) (500.0, -10.586193845483507) (520.0, -9.932747055012985) (540.0, -9.656510617517299) (560.0, -9.568848961177027) (580.0, -9.443834360449639) (600.0, -9.377096240614472) (620.0, -9.34985177526382) (640.0, -9.219731247974584) (660.0, -9.010544310582349) (680.0, -5.4)};

\addplot [blue!25] fill between [ of=B and A ];

\end{axis}
\end{tikzpicture}

%% file: sections/7-evaluation/setting.tex
\subsection{Setting}
\label{sec:setting}
As a testbed for attacks, we leveraged our cyber range~\cite{Russo2020} that is integrated with the Shil~\cite{DAgostino2020} infrastructure.
It emulates a realistic ship navigation network, device sensors, and equipment as detailed in~\ref{sec:ship-navigation-network}.
In particular, it hosts our extension of the Bridge Command (BC)~\cite{Packer22} ship and radar simulator that implements an add-on for transmitting radar data using the ASTERIX CAT-240 protocol.
Thus, we simulated sensors devices by transmitting data using NMEA and a radar antenna with an accuracy compatible with the performance standards, i.e., a bearing resolution of \ang{1} and a range resolution of 10.85 meters in the range scale of 12 nautical miles.
We used a digital PPI produced by a leading manufacturer and widely adopted in naval and commercial ships for displaying the video data and tracking targets with ARPA.
Finally, we connected two Debian GNU/Linux 11 virtual machines hosted by VMWare ESXi 7.0U3 and configured with 1 Intel Xeon Gold 6252N at 2.3GHz, 4GB of RAM, and 30GB of storage. 
The first acts as a bridge workstation and runs a Proof-of-Concept (PoC) implementation of the malware.
The PoC has been developed in Rust, amounts to 3761 lines of code, and supports cross-compiling to different architecture, e.g., x86 and ARM, and operating systems, e.g., Windows, Linux, and OS X.
The version we used is a Linux executable file with a size of $1171$KiB.  
Finally, the second virtual machine hosts our detection system.

%% file: sections/7-evaluation/results.tex
\subsection{Results}
We generated on the BC simulator 25 instances of three scenarios that set the environment for executing the attacks we presented in~ Section~\ref{sec:attackimpl}.
Assuming that $\mathcal{U}$ is the uniform random distribution, each instance features a number $\mathcal{U}_{\{2,8\}}$ of ships.
We placed them at a distance of $\mathcal{U}_{[3.5,5]}$ nautical miles and $\pm U_{[10,80]}$ degrees w.r.t. the bow of the victim, and moving them at a random speed of $\mathcal{U}_{[2,12]}$ knots.
For the Ghost Ship attack, we also added the ghost ship that moved at a speed of 10 knots and with the trajectory depicted in Figure~\ref{fig:ghost-ship}. 
For the Ship Trajectory Hijack attack, we added a ship with a speed of $\mathcal{U}_{[2.5,5.0]}$ knots to create an overtaking situation with the victim.
The victim ship moved at a speed of 10 knots, and all the vessels kept their course and speed constant.

The radar under test received ASTERIX data from the BC and tracked the surrounding vessels using ARPA during the test execution.
We set our radar TCPA default alarm (see Section~\ref{sec:background-arpa}) at $15$ minutes.

Each experiment lasted 60 seconds for the DoS, 120 seconds for the Ghost Ship, and 600 seconds for the Ship Trajectory Hijack. 

At the same time, we run our detection solution configured with six policies.
We divided them into two groups, namely \emph {categorical} or \emph{statistical}.

A categorical policy detects if a given field assumes a specific value with a probability greater than $0.99$.
After it activates, the generated policy enforces the above value when it exists.
We configured categorical policies for the $center\_bias$ ($P_1$) and $n\_cells$ ($P_2$) fields.

A statistical policy verifies if it can construct an estimator for a field value. 
After it activates, the generated policy tests if the given field is consistent with the null hypothesis on the constructed estimator.
We configured statistical policies for $(i)$ the azimuthal span (see Example~\ref{ex:azspan-policy}), i.e., enforcing the rotation speed to be constant in between packets ($P_3$), $(ii)$ the monotonicity of the $message\_id$ field ($P_4$), $(iii)$ the number of entries belonging to each aggregation, i.e., enforcing the rotation speed to be constant within a revolution ($P_5)$, and $(iv)$ the number of aggregation in the historian within a fixed time period, i.e., enforcing the rotation speed to be constant across revolutions ($P_6$).

For each experiment, we recorded performance figures of the malware (see Table~\ref{tab:attack-performance}) and the detection system (see Table~\ref{tab:detection-performance}).
We report the CPU and RAM usage for both attack and defense.
In Table~\ref{tab:attack-performance}, we also include statistics about ASTERIX traffic and the percentage of the malicious traffic compared to the legitimate traffic.

Then, we used information emitted by the ARPA system to evaluate the effectiveness of attacks.

For DoS attacks, we considered each experiment as a success if the radar image corruption caused at least one of the following two impacts on the ARPA system: $(i)$ it lost tracking of a target, or $(ii)$ returned nonphysical data about targets (e.g., a speed $\ge 100 [m/s]$, or an acceleration at a rate $\ge 10 [m/s^2]$). 
Results showed that DoS attacks had a success rate of 100\%.

For PPI poisoning attacks, we considered both the accuracy of the malware w.r.t. the trajectory to reproduce and the success conditions.
We measured for each \texttt{TTM} the absolute error between the desired courses and speeds and the ones emitted by the ARPA system to estimate the accuracy.
In 29 out of 50 (58\%) cases, the course did not deviate by more than $1^\circ$, with every trajectory within $10^\circ$.
In 40 out of 50 (80\%) cases, the speed did not deviate by more than $0.1$ knots, with every speed within $0.5$ knots.

To assess the success of each experiment, we considered the TCPA for the ghost and hijacked ships.
In Figure~\ref{fig:attack-tcpa-behavior}, we present the evolution of the TCPA value during all the experiments by highlighting the complete range of the distribution and the average trend.
Ghost Ship attacks required as a successful result the TCPA of the ghost ship to decrease to the collision alert threshold.
On the contrary, Trajectory Hijacking attacks required the TCPA of the overtaken ship to grow up to indicate an increasing trend, i.e., a negative value.
Figure~\ref{fig:attack-tcpa-behavior} highlights that the two attacks had a success rate of 100\% since the TCPA always complied with the expected trend.

Finally, we considered the accuracy of our detection system.
In Table~\ref{tab:detection_res}, we summarize the total packets for each attack by identifying them as legitimate or malicious and how our detection system classified them in terms of true or false positives. 
In Table~\ref{tab:matched-policies}, we outline the policies that each attack triggered during the experiments by considering the percentage of malicious packets that they matched.

%% file: sections/7-evaluation/discussion.tex
\subsection{Discussion}
\label{sec:discussion}

Concerning the techniques and requirements introduced in Section~\ref{sec:attack-techniques}, we put forward the following considerations related to our malware and the feasibility of the  attacks.
First, the malware is a small and cross-compilable executable and can successfully carry out coarse-grained and fine-grained attacks using low computational and memory resources.
These features allow an adversary to install it and execute the presented attacks on a wide range of INS configurations, including legacy and embedded systems.
The experimental results indicate that the malware has a very limited footprint on both the usage of computational/memory resources and the amount malicious traffic w.r.t. the legitimate one. This fact confirms that the malware can act stealthy and execute PPI poisoning attacks without causing noticeable effects.
For DoS attacks, we showed that it could leverage features of the ASTERIX protocol to obtain the above small footprint.
Moreover, overhearing NMEA traffic allows the malware to acquire the information and requirements to operate independently and without communicating outside the INS.

Regarding attacks, the experiments executed in our cyber range with multi-ships scenarios and the commercial PPI proved their feasibility. Results also showed that the attacks could achieve a high degree of realism by precisely simulating the behavior of a vessel on a predetermined trajectory.
Such realism and the capability of these attacks to inject AIS traffic for cheating the cross-checking with INS equipment show the high deception capability against maritime operators and the potential to cause catastrophic impacts.

Regarding the detection system, we show that policies enforcing the performance standards for an antenna ($P_3$, $P_5$, $P_6$), ASTERIX protocol specifications ($P_4$), and the expected behavior inferred from the on-board configurations ($P_1$, $P_2$) enabled the detection of all the attacks with high accuracy.
The resulting performance highlighted that our system requires minimal resource footprint.
Finally, it is worth noting that the detection operates only on packets header and can ensure similar performances on other antenna types, even with higher video resolutions.

%% file: sections/8-conclusion/section.tex
\section{Conclusion}
\label{sec:conclusion}
In this paper, we identified that configurations and standard protocols commonly used in ships and related to INSs are vulnerable to novel attacks targeted at the maritime radar systems.
We demonstrated how a suitably equipped attacker could inject a targeted malware leveraging the specific technological environment to autonomously execute the attacks.

Radar is an essential aid to ensure safe navigation, and the consequences of these attacks are significant.
We showed that they could lead to a high-impact disruption of normal operativity up to stealthy alterations causing awareness mismatches between the victim and other ships nearby and increasing the potential for hazardous situations.

We also developed a detection system able to recognize such attacks with a high level of accuracy.
The distinguishing features of our proposal are $(i)$ the self-adaptation to each onboard configuration, $(ii)$ the modeling of regulatory and expected behavior to identify known and unknown attacks, $(iii)$ the possibility of running it without altering onboard systems, and $(iv)$ the minimal resource footprint.

Future directions include proposing training activities on our cyber range to improve the awareness of the maritime operators also in response to these new types of attacks.